\documentclass[aps,twocolumn,10pt,amsmath,amssymb,nofootinbib,superscriptaddress,showkeys,showpacs,pra]{revtex4-1}

\usepackage{epsfig} 
\usepackage{amsmath} 
\usepackage{pifont}
\usepackage{graphicx}
\usepackage{color}
\usepackage{float}
\usepackage{soul,xcolor}
\usepackage[export]{adjustbox}
\usepackage{multirow}
\usepackage[utf8]{inputenc}
\usepackage{tabularx,colortbl}

\usepackage{bbold}
\usepackage{braket}
\usepackage{titlesec}
\usepackage{placeins}
\usepackage{epstopdf}

\DeclareMathOperator*{\dprime}{\prime \prime}

\DeclareMathOperator*{\avgA}{\langle \hat{A} \rangle}
\DeclareMathOperator*{\avgB}{\langle \hat{B} \rangle}
\DeclareMathOperator*{\avgc}{\langle \hat{C} \rangle}

\DeclareMathOperator*{\avgAA}{\langle \hat{A}^2 \rangle}
\DeclareMathOperator*{\avgBB}{\langle \hat{B}^2 \rangle}
\DeclareMathOperator*{\avgcc}{\langle \hat{C}^2 \rangle}

\DeclareMathOperator*{\avgAdgr}{\langle \hat{A}^\dagger \rangle}
\DeclareMathOperator*{\avgBdgr}{\langle \hat{B}^\dagger \rangle}
\DeclareMathOperator*{\avgcdgr}{\langle \hat{C}^\dagger \rangle}

\DeclareMathOperator*{\avgAdgrAdgr}{\langle (\hat{A}^{\dagger})^2 \rangle}
\DeclareMathOperator*{\avgBdgrBdgr}{\langle (\hat{B}^{\dagger})^2 \rangle}
\DeclareMathOperator*{\avgcdgrcdgr}{\langle (\hat{C}^{\dagger})^2 \rangle}

\DeclareMathOperator*{\avgAdgrA}{\langle \hat{A}^\dagger \hat{A} \rangle}
\DeclareMathOperator*{\avgBdgrB}{\langle \hat{B}^\dagger \hat{B} \rangle}
\DeclareMathOperator*{\avgcdgrc}{\langle \hat{C}^\dagger \hat{C} \rangle}

\DeclareMathOperator*{\avgAB}{\langle \hat{A} \hat{B} \rangle}
\DeclareMathOperator*{\avgABdgr}{\langle \hat{A} \hat{B}^\dagger \rangle}
\DeclareMathOperator*{\avgAdgrB}{\langle \hat{A}^\dagger \hat{B} \rangle}
\DeclareMathOperator*{\avgAdgrBdgr}{\langle \hat{A}^\dagger \hat{B}^\dagger \rangle}

\DeclareMathOperator*{\avgBc}{\langle \hat{B} \hat{C} \rangle}
\DeclareMathOperator*{\avgBcdgr}{\langle \hat{B} \hat{C}^\dagger \rangle}
\DeclareMathOperator*{\avgBdgrc}{\langle \hat{B}^\dagger \hat{C} \rangle}
\DeclareMathOperator*{\avgBdgrcdgr}{\langle \hat{B}^\dagger \hat{C}^\dagger \rangle}

\DeclareMathOperator*{\avgAc}{\langle \hat{A} \hat{C} \rangle}
\DeclareMathOperator*{\avgAcdgr}{\langle \hat{A} \hat{C}^\dagger \rangle}
\DeclareMathOperator*{\avgAdgrc}{\langle \hat{A}^\dagger \hat{C} \rangle}
\DeclareMathOperator*{\avgAdgrcdgr}{\langle \hat{A}^\dagger \hat{C}^\dagger \rangle}

\DeclareMathOperator*{\AT}{\overset{\text{\ding{51}}}{A}}
\DeclareMathOperator*{\BT}{\overset{\text{\ding{51}}}{B}}
\DeclareMathOperator*{\CT}{\overset{\text{\ding{51}}}{C}}
\DeclareMathOperator*{\ABT}{\overset{\text{\ding{51}}}{AB}}
\DeclareMathOperator*{\BCT}{\overset{\text{\ding{51}}}{BC}}
\DeclareMathOperator*{\ACT}{\overset{\text{\ding{51}}}{AC}}
\DeclareMathOperator*{\BAT}{\overset{\text{\ding{51}}}{BA}}
\DeclareMathOperator*{\CBT}{\overset{\text{\ding{51}}}{CB}}
\DeclareMathOperator*{\CAT}{\overset{\text{\ding{51}}}{CA}}

\DeclareMathOperator*{\ABF}{\overset{\text{\ding{55}}}{AB}}

\DeclareMathOperator*{\ACF}{\overset{\text{\ding{55}}}{AC}}
\DeclareMathOperator*{\BAF}{\overset{\text{\ding{55}}}{BA}}

\DeclareMathOperator*{\CAF}{\overset{\text{\ding{55}}}{CA}}

\DeclareMathOperator*{\ABcT}{\overset{\text{\ding{51}}}{AB|C}}
\DeclareMathOperator*{\BCaT}{\overset{\text{\ding{51}}}{BC|A}}
\DeclareMathOperator*{\ACbT}{\overset{\text{\ding{51}}}{AC|B}}

\DeclareMathOperator*{\ABcF}{\overset{\text{\ding{55}}}{AB|C}}
\DeclareMathOperator*{\BCaF}{\overset{\text{\ding{55}}}{BC|A}}
\DeclareMathOperator*{\ACbF}{\overset{\text{\ding{55}}}{AC|B}}

\begin{document}	
	
		\title{Probing non-classicality in an optically-driven  cavity with two atomic ensembles}
	\author{Javid Naikoo}
	\email{naikoo.1@iitj.ac.in}
	\affiliation{Indian Institute of Technology Jodhpur, Jodhpur 342011, India}
	
	\author{Kishore Thapliyal}
	\email{tkishore36@yahoo.com}
	\affiliation{Jaypee Institute of Information Technology, A-10, Sector-62, Noida UP-201307, India}
	
	\author{Anirban Pathak}
	\email{anirban.pathak@jiit.ac.in}
	\affiliation{Jaypee Institute of Information Technology, A-10, Sector-62, Noida UP-201307, India}
	
	\author{Subhashish Banerjee}
	\email{subhashish@iitj.ac.in}
	\affiliation{Indian Institute of Technology Jodhpur, Jodhpur 342011, India}




\begin{abstract}
The possibility of observing non-classical features in a physical system comprised of a cavity with two ensembles of two-level atoms has been investigated by considering  different configurations of the ensembles with respect to the Node and Antinode of the cavity field under the framework of open quantum systems. The study reveals the strong presence of  non-classical characters in the physical system by establishing the existence of many facets of non-classicality, such as the sub-Poissonian boson statistics and squeezing in single modes, intermodal squeezing,  intermodal entanglement, antibunching, and steering. The effect of a number of parameters, characterizing the physical system, on the different aspects of non-classicality is also investigated. Specifically, it is observed that the depth of the non-classicality witnessing parameters can be enhanced by externally driving one of the ensembles with an optical field. The work is done in the presence of open system effects, in particular, use is made of Langevin equations along with a  suitable perturbative technique.  
\end{abstract}
\maketitle 
	
	\section{Introduction} \label{Introduction}
 Quantum mechanics has emerged as the best known model of nature. Thanks to the spectacular success achieved over the last hundred years. However, only in the last few decades, it is understood that quantum mechanics can even be used to design devices that can outperform their classical counterparts. This quantum power of devices is obtained by exploiting non-classical states, i.e., states having no classical analogue and more technically, the quantum states having negative values of Glauber-Sudarshan $P$-function \cite{glauber1963coherent,sudarshan1963equivalence}. Such states are not rare in nature, and entangled and steering states \cite{reid1989demonstration}, squeezed states \cite{loudon1987squeezed}, antibunched states \cite{teich1983antibunching} are typical examples of non-classical states. The existence of such states were known (at least theoretically) since a long time. In fact, squeezing \cite{kennard1927quantenmechanik}, entanglement \cite{einstein1935can}, and steering \cite{schrodinger1935discussion} were studied even before the pioneering work of Sudarshan \cite{sudarshan1963equivalence} that provided  a necessary and sufficient criterion of non-classicality in terms of negativity of $P$-function. However,   
various interesting applications of these non-classical states were realized only recently with the advent of quantum information processing \cite{hillery2000quantum,furusawa1998unconditional,yuan2002electrically,ekert1991quantum,bennett1993teleporting,bennett1992communication} and various facets of atom optics and quantum optics \cite{agarwal2013quantum,scully1997quantum}. For example, squeezed vacuum state has been used successfully in detecting gravitational waves in the well known LIGO experiment \cite{abbott2016observation,abbott2016gw151226}; squeezed states are also used in continuous variable quantum secure and insecure communication \cite{hillery2000quantum,furusawa1998unconditional}; entanglement is established to be useful in both continuous and discrete variable quantum cryptography \cite{ekert1991quantum,hillery2000quantum}, and in the realization of schemes for teleportation \cite{bennett1993teleporting} and 
dense coding \cite{bennett1992communication}. Additionally, the steerable states provide one-side device independent quantum cryptography \cite{branciard2012one}. Furthermore, powerful quantum algorithms for unsorted database search \cite{grover1997quantum} to factorization \cite{shor1999polynomial}, discrete logarithm problem \cite{shor1999polynomial} to machine learning \cite{biamonte2017quantum} have repeatedly established that quantum computers (which naturally use non-classical states) can outperform classical computers. In brief, in the last few years, on one hand, we have seen various applications of non-classical states, and on the other hand, non-classical features have been reported in a variety of physical systems \cite{thapliyal2015quasiprobability,sen2013intermodal,giri2014single,alam2017lower}, including but not restricted to two-mode Bose-Einstein condensates \cite{giri2014single,giri2017nonclassicality}, optical couplers \cite{thapliyal2014higher,thapliyal2014nonclassical}, optomechanical \cite{alam2017lower,bose1997preparation} and optomechanics-like systems \cite{zhang2012role,alam2017lower}, atoms and quantum dot in a cavity \cite{baghshahi2015generation,majumdar2012probing}. Many of these systems involve different types of cavity which can be produced and manipulated experimentally \cite{slusher1985observation,mundt2002coupling,birnbaum2005theory}. Naturally, interest in such systems has been considerably enhanced in the recent past. Apart from the applicability of the non-classical states, and the possibilities of generation and manipulation of these states, another interesting factor that has enhanced the interest on the non-classical features present in these systems, is the fact that in contrast to the traditional view that  quantum mechanics is the science of the microscopic world, these systems having non-classical properties are often macroscopic \cite{aspelmeyer2014cavity}. 

Above facts have motivated us to study non-classical features of a particular macroscopic system shown in Fig. \ref{Figure}. To be specific, in this paper, we aim to investigate the possibility of observing signatures of various types of non-classicality in a physical system comprised of a cavity with two ensembles of two-level atoms,  placed in different configurations with respect to the Node and Antinode of the cavity field, like, Antinode-Antinode (AA), Antinode-Node (AN), Node-Antinode (NA) and Node-Node (NN). To clearly visualize these configurations, we may note that in AN configuration, one of the ensembles  is placed in the Node  position of the cavity field and the other one is placed in the Antinode position of the cavity field. Similarly, one can visualize the other configurations studied in this paper. Previously, this system was used to study  electromagnetically-induced-transparency-like (EIT-like) phenomenon in \cite{turek2013electromagnetically}, where an EIT-like phenomenon was observed to appear (disappear) for the NA (AN) configuration. In what follows, we will study the possibilities of observing various types of single mode (e.g., squeezing, sub-Poissonian boson statistics) and intermodal non-classicality (e.g., intermodal squeezing, antibunching, two and three mode entanglement, steering) in this system by  considering that one of the atomic ensembles is driven by an external optical field, and will establish that this external field can be used to control the amount of non-classicality. 

We have already noted that the negativity of $P$-function provides us a necessary and sufficient criterion of non-classicality. However, the $P$-function is not always well behaved, and there does not exist any general procedure that can be adapted to experimentally measure it. As a consequence, a set of operational criteria for non-classicality have been developed over the years. The majority of these non-classical criteria (in fact, all the criteria used in this work) do not provide any quantitative measure of non-classicality\footnote{Of course, there are some measure of non-classicality, but each of them have some issues \cite{miranowicz2015statistical}, and we have not used any of them in the present study.} and they are only sufficient criteria. In fact, there exists an infinite set of non-classicality criteria involving moments of annihilation and creation operators that are equivalent to the $P$-function, but any finite subset of that would be sufficient only. In this paper, we have used a few  such moment-based criteria of non-classicality \cite{richter2002nonclassicality,miranowicz2010testing}, each of which is a sufficient criterion only. As none of these criteria provides any quantitative measure of non-classicality (i.e., as they only provide signatures of non-classicality), in what follows, these sufficient criteria are frequently referred to as witnesses of non-classicality. In what follows, through these criteria, different features of non-classicality are witnessed under the influence of open quantum system evolution.

The effect of the ambient environment is a permanent fixture of nature and needs to be taken into account, especially in experiments related to non-classical features which are  known to be influenced appreciably by the environmental effects. As the present work aims to reveal the non-classical features present in the system of interest, it would be apt to consider the effect of the environment in our calculations. Such effects are taken into account systematically by using the framework of open quantum systems \cite{breuer2002theory}. Specifically, decoherence and dissipation are well known open system effects \cite{banerjeedecdiss} and have been studied on  myriad aspects of quantum information, such as in holonomic quantum computation \cite{banerjee2008geometric}, environmental deletion \cite{srikanth2007environment}, noisy quantum walks \cite{chandrashekar2007symmetries}, quantum cryptography \cite{srinatha2014quantum} and the effect of squeezing on channel capacity \cite{srikanth2008squeezed}. A precursor of the present theme of non-classical correlations in the presence of open system effects can be found in \cite{chakrabarty2010study}. Here, we adapt open system effects on our system of interest by using the formalism of Langevin equations, which is basically the stochastic equations of motion approach \cite{scully1997quantum}. Specifically, the equations of motion for each system mode in the Heisenberg picture are obtained by eliminating the environmental degrees of freedom. The obtained equations of motion for different system modes are usually coupled differential equations and are solved using various peturbative techniques. Here, we have used a perturbative technique that approximates all the higher-order correlations in terms of second-order correlations \cite{anglin2001dynamics}. The technique has been recently used to study non-classicality in Raman amplifier \cite{ooi2007correlation} and optomechanical oscillator \cite{singh2014quantum}.

The rest of the paper is organized as follows.  In Section \ref{Model}, we describe the model used in this work in the context of open quantum systems. Section \ref{Criteria}  gives a brief introduction to the various witnesses of non-classicality used in this work. Subsequently, in Section \ref{Results}, we present temporal variation of various witnesses of non-classicality and discuss the significance of the results obtained in this work. Finally, the paper is concluded in Section \ref{Conclusion}.

 \section{Cavity containing  two ensembles of two-level atoms} \label{Model}
 
\begin{figure}[ht] 
 	\centering
 	\begin{tabular}{cc}
 		\includegraphics[width=80mm]{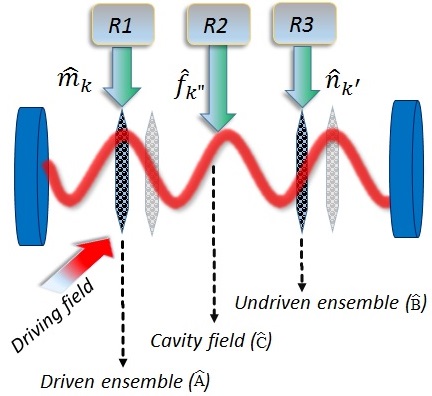}
 	\end{tabular}
 	
 	\caption{(Color online) The schematic representation of the model consisting of a cavity embedded with two ensembles of two-level atoms. The left ensemble $\mathcal{S}_A$, with the excitation mode $\hat{A}$, is driven by an external field of frequency $\omega_f$. The system is studied in configurations: Antinode-Antinode (AA), Antinode-Node (AN), Node-Antinode (NA) and Node-Node (NN). The left ensemble (the driven ensemble), the right ensemble (the undriven ensemble), and the cavity field interact with their independent reservoir modes represented by corresponding annihilation operators $\hat{m}$, $\hat{n}$, and $\hat{f}$, respectively.  }
 	\label{Figure}
 \end{figure}

 The physical system of our interest is briefly described in the previous section and it is schematically shown in Fig. \ref{Figure}. In this section, we wish to describe the system in more detail. To begin with, we note that the model physical system of our interest is considered to be made of a single mode cavity ($\mathcal{S}_{\rm cavity}$) which contains two ensembles ($\mathcal{S}_A-$left ensemble and $\mathcal{S}_B-$right ensemble) of two level atoms \cite{turek2013electromagnetically}. The left ensemble is driven by a classical optical field having frequency $\omega_f$.

	The Hamiltonian for the total  system $S\equiv \mathcal{S}_A + \mathcal{S}_B + \mathcal{S}_{\rm cavity}$,  can be expressed in terms of collective excitation operators $\hat{A}$ and $\hat{B}$ in the following form \cite{turek2013electromagnetically}:
	
	\begin{align}
	\hat{H} &= \omega_c \hat{C}^{\dagger}\hat{C} + \omega_a  \nonumber \hat{A}^{\dagger}\hat{A} + \omega_b \hat{B}^{\dagger} \hat{B} +\left\{ G_A \hat{C}\hat{A}^{\dagger} + G_B \hat{C} \hat{B}^{\dagger}\right. \\&+ \left.\chi \hat{A}^{\dagger} e^{-i\omega_f t}	+ \rm{H.c.} \right\}, \label{H1}
	\end{align} 
	with
	\begin{align*}
	\hat{A} = \frac{1}{\sqrt{N_A}} \sum\limits_{i=1}^{N_A} \sigma_{+,A}^{i} ~~~~\text{and}
	\quad    
\hat{B} = \frac{1}{\sqrt{N_B}} \sum\limits_{j=1}^{N_B} \sigma_{+,B}^{j},
	\end{align*}
	where  H.c., stands for Hermitian conjugate, and $\sigma_{+,x}^l = \ket{e_{x}^{(l)}}\bra{g_{x}^{(l)}}$ and $\sigma_{-,x}^l = \ket{g_{x}^{(l)}}\bra{e_{x}^{(l)}}$ are the quasispin operators for the $l$-th atom in ensemble $\mathcal{S}_x$ ($x \in \{A,B\}$). The operator $\hat{C}\, (\hat{C}^{\dagger})$ represents the annihilation (creation) operator for the cavity mode. Also, $G_A = g_A \sqrt{N_A}$, $G_B = g_B \sqrt{N_B}$ and $\chi = \Omega \sqrt{N_A}$, where $g_A$ ($g_B$) is the strength with which the atoms in left (right) ensemble couple with the cavity field. Similarly, $\Omega$ (or equivalently $\chi$) corresponds to the coupling strength between the atoms in the driven ensemble and the driving field. In the limit of low excitation and large number of atoms ($N_A$ and $N_B$), the operators $\hat{A}$ and $\hat{B}$ satisfy the bosonic commutation relations, i.e., $[\hat{A}, \hat{A}^\dagger] \approx [\hat{B}, \hat{B}^\dagger] \approx \mathbb{1}$, and  $[\hat{A}, \hat{B}] \approx [\hat{A}, \hat{B}^\dagger] \approx 0$. Therefore, under these conditions, $\hat{A}$ and $\hat{B}$ can be treated as the annihilation operators for the collective \textit{excitation modes} corresponding to ensembles $\mathcal{S}_A$ and $\mathcal{S}_B$, respectively \cite{turek2013electromagnetically}. \\

	  In the interaction picture, the Hamiltonian  given by Eq. (\ref{H1}) can be expressed in terms of the photonic cavity mode $\hat{C}$ and the ensemble excitation modes $\hat{A}$ and $\hat{B}$, in the following simplified form \cite{turek2013electromagnetically}
	\begin{align}
    	\hat{H}_{S} &= \Delta_c \hat{C}^{\dagger}\hat{C} + \Delta_a  \nonumber \hat{A}^{\dagger}\hat{A} + \Delta_b \hat{B}^{\dagger} \hat{B} + \{G_A \hat{C}\hat{A}^{\dagger} + G_B \hat{C} \hat{B}^{\dagger} \\&+ \chi \hat{A}^{\dagger} 	+ \rm{H.c.}\},
	\end{align}
where $\Delta_r = \omega_f - \omega_r$ ($r\in \{a,b,c\}$) represents  the frequency detuning of the driven ensemble frequency ($\omega_a$), the un-driven ensemble frequency ($\omega_b$) and the cavity field frequency ($\omega_c$),  with respect to the driving frequency $\omega_f$.

Open quantum system effects are now taken into consideration by allowing the interaction of the photonic  cavity mode $\hat{C}$ and the collective excitation modes of the ensembles (i.e., modes $\hat{A}$ and $\hat{B}$) with their respective reservoirs. This interaction is modeled, under the Markovian white noise approximation, by coupling  each mode to a reservoir made up of a collection of harmonic oscillators \cite{breuer2002theory}. As a result, the total Hamiltonian is modified to
\begin{equation}
\hat{H} = \hat{H_S} + \hat{H_R} + \hat{H}_{SR},
\end{equation} 
such that,
\begin{align}
\hat{H}_R &= \sum\limits_{k} \omega_k \hat{m}_k^\dagger \hat{m}_k 
+ \sum\limits_{k^{\prime}} \omega_{k^{\prime}} \hat{n}_{k^{\prime}}^\dagger \hat{n}_{k^{\prime}} 
+ \sum\limits_{k^{\dprime}} \omega_{k^{\dprime}} \hat{f}_{k^{\dprime}}^\dagger \hat{f}_{k^{\dprime}},
\end{align}

\begin{align}
\hat{H}_{SR} &= \sum\limits_{k} g_k(\hat{m}_k^\dagger \hat{A} + \hat{A}^\dagger \hat{m}_k)    \nonumber
+ \sum\limits_{k^{\prime}} g_{k^{\prime}}(\hat{n}_{k{^\prime}}^{\dagger} \hat{B} + \hat{B}^{\dagger} \hat{n}_{k^{\prime}}) 
\\& +\sum\limits_{k^{\dprime}} g_{k^{\dprime}}(\hat{f}_{k^{\dprime}}^{\dagger} \hat{C} + \hat{C}^{\dagger} \hat{f}_{k^{\dprime}}),
\end{align}
where $\hat{m}(\hat{m}^\dagger)$, $\hat{n}(\hat{n}^\dagger)$ and $\hat{f}(\hat{f}^\dagger)$ are the annihilation (creation) operators corresponding to the reservoirs which interact with and damp the driven ensemble mode $\hat{A}$, the un-driven ensemble mode $\hat{B}$ and the cavity mode $\hat{C}$, respectively. Here, and in what follows, $S$ and $R$ in the subscript correspond to the system and reservoir, respectively. The resulting (Langevin) equations, for the system operators should include, in addition to the damping terms, the noise operators which would produce fluctuations \cite{scully1997quantum}.
We can now explicitly write the Langevin equations for the cavity and atomic ensemble modes. Specifically, the Langevin equations for the cavity mode can be written as
    \begin{equation}
   \frac{d \hat{C}}{dt} = -i\Delta_c \hat{C} - i G_A \hat{A} - i G_B \hat{B} -\frac{\Gamma_c}{2} \hat{C} + \hat{F}_c, \label{Copt}
    \end{equation}
    where $\Gamma_c$ is the decay constant and $\hat{F}_c$ is the noise operator. For the initially uncorrelated subsystems, the initial density matrix can be considered as separable and thus in the tensor product form $\rho = \rho_S \otimes \rho_R$, and similarly it may be considered that the expectation value of operator $M = M_S \otimes M_R$ factors as $\langle M_S \rangle \langle M_R \rangle$. Eq. (\ref{Copt}) is an operator equation and it is not easy to obtain an analytic solution of this type of equations. Keeping this in mind, here we adapt a strategy used in Refs. \cite{anglin2001dynamics,ooi2007correlation,singh2014quantum}. Following this strategy, we begin our solution scheme by taking an average of each term appearing in this equation with respect to the state $\rho$. This step yields a differential equation of the average of $\hat{C}$ in terms of averages of $\hat{A}$ and $\hat{B}$. Note that this step transforms the operator differential equation into a $c$-number differential equation which is much easier to handle. Assuming each  reservoir to be in thermal equilibrium at temperature T, we can average over the system and reservoir degrees of freedom and using the fact that the reservoir average of the noise operator vanishes $\langle \hat{F}_c \rangle_{R} = 0$ \cite{scully1997quantum}, we end up with the following equation of motion 
	\begin{equation}
	\frac{d \avgc}{dt} = -i\Delta_c   \avgc -i G_A \avgA -iG_B \avgB - \frac{\Gamma_c}{2} \avgc .\label{eq:Lang-c}
	\end{equation}
	Similarly, we can  obtain the Langevin equations for modes $\hat{A}$ and $\hat{B}$, and for all the second order terms in creation  and annihilation operators. Averaging each term present in these operator differential equations would lead to  a set of coupled ordinary differential equations involving various statistical quantities of interest. In general, these coupled differential equations are required to be decoupled using an appropriate approximation scheme. To maintain the flow of the paper, we have reported this set of equations in Appendix ${\bf A}$, where we have also described the method adopted in this paper to decouple (solve) them. Now, we may move to the next section, where we will briefly describe various measurable criteria of non-classicality which will be used in the subsequent section to investigate the presence of non-classicality in the physical system of our interest.

\section{ Criteria of non-classicality} \label{Criteria}

Nonclassicality is a multifaceted entity. It is an important problem to understand various aspects of non-classicality in the context of open quantum systems \cite{chakrabarty2010study}. 
From the perspective of quantum optics there are different witnesses of non-classicality of the radiation field. For example, the Mandel parameter $Q_M < 0$, gives a sufficient condition for the field to be non-classical \cite{agarwal2013quantum};   single and multimode squeezing conditions reveal the non-classical character of a state arising due to the field fluctuation \cite{loudon1987squeezed}; Hillery-Zubairy criteria provide sufficient conditions in the form of a family of inequalities for detecting entanglement\cite{hillery2006entanglement}. These criteria can be casted in terms of the bosonic \(creation\) and \(annihilation\) operators as discussed below. Thus, analysis of the various types of non-classicality in the context of above developed model characterized through the witnesses listed here, can be carried out.
\begin{itemize}
	\item \textit{The Mandel  $Q_M$ parameter}: Defined as the normalized variance of the boson distribution, this measure characterizes the non-classicality of a radiation field in the context of the photon number distribution.   Quantitatively,
	\begin{equation}
	Q_M = \frac{\langle(a^\dagger a)^2 \rangle - \langle a^\dagger a \rangle^2 - \langle a^\dagger a  \rangle }{\langle a^\dagger a  \rangle}  \label{mandel parameter}.
	\end{equation}

Since the minimum value of \(\langle(a^\dagger a)^2 \rangle - \langle a^\dagger a \rangle^2\) is zero, the Mandel parameter has a lower bound of \(-1\), and it provides the criterion for observing different photon statistics as follows:
\begin{equation}
Q_M  \begin{cases}
< 0 &  {\rm sub-Poissonian~ field},\\
=0 &  {\rm coherent~  (Poissonian)~ field}, \\
> 0 & {\rm super-Poissonian~ field}.
\end{cases} 
\end{equation}
 \item  \textit{Antibunching}: A closely related phenomena is photon antibunching, given usually in terms of the two-time light intensity correlation function \cite{mandel1995optical}, $g^{(2)}(\tau) = \langle n_1(t)n_2(t + \tau) \rangle / \langle n_1(t) \rangle \langle n_2(t + \tau) \rangle$, where $n_i(t)$ is the number of counts registered on $i$th detector at time $t$. A quantum state is referred to as an antibunched if \(g^{(2)}(0)<g^{(2)}(\tau) \). Interestingly, it was shown in the past to be closely related to the Mandel parameter \cite{fox2006quantum}. The correlation \(g^{(2)}(0) \) characterizes the antibunched, the coherent and the bunched fields as:
 \begin{equation}
 g^{(2)}(0)  \begin{cases}
 < 1 &  \rm antibunched,\\
 =1 &  \rm coherent, \\
 > 1 & \rm bunched.
 \end{cases} 
 \end{equation}
  Therefore, for a single field with annihilation operator $a$, the criterion for antibunching can also be written as  \cite{pathak2006control}
 \begin{equation} \label{SingleAntiBun}
\mathcal{A}_{a}  = \langle a^{\dagger 2} a^2 \rangle - \langle a^\dagger a \rangle^2 < 0,
 \end{equation}
 i.e., the negative values of Mandel parameter also establish antibunching.
 Further, the intermodal antibunching is witnessed by using the following criterion \cite{thapliyal2014nonclassical}
 \begin{equation} \label{InterModalAntiBun}
 \mathcal{A}_{ab}  = \langle a^{\dagger} b^{\dagger} b a \rangle - \langle a^{\dagger} a \rangle \langle b^{\dagger} b \rangle < 0.
 \end{equation}
  
\item \textit{Squeezing}: This measure delineates the non-classicality of a field in the context of the fluctuations in the quadratures $X_a$ and $Y_a$ of the field (with \(annihilation\) operator \(a\)), defined as   
\begin{equation}
X_a = \frac{a + a^\dagger}{2}  \qquad  Y_a = \frac{a - a^\dagger}{2i}  \label{single_quadrature}.
\end{equation}
 The criteria for the non-classical signature  in the field is given, in terms of the variances in the quadratures, as follows \cite{loudon1987squeezed}
 \begin{eqnarray} 
\langle X^2_a \rangle - \langle X_a \rangle^2 = (\Delta X_a)^2 < \frac{1}{4} \label{SMSX} 
\end{eqnarray}
or
\begin{eqnarray} 
\langle Y^2_a \rangle - \langle Y_a \rangle^2 = (\Delta Y_a)^2 < \frac{1}{4}. \label{SMSY} 
 \end{eqnarray}
 We can also define the intermodal quadrature operators \(X_{ab} = (a + a^\dagger + b + b^\dagger)/2\sqrt{2}\) and \(Y_{ab} = (a - a^\dagger + b - b^\dagger)/2i\sqrt{2}\), such that the intermodal squeezing criterion is given by 
 
\begin{subequations}
 \begin{eqnarray}
(\Delta X_{ab})^2 < \frac{1}{4} \label{IMS1}\\
{\rm or}\nonumber\\
(\Delta Y_{ab})^2 < \frac{1}{4}. \label{IMS2}
 \end{eqnarray}\label{IMS}
\end{subequations}

  \item \textit{Duan el al.'s~criterion of entanglement}:  For two  systems \(A\) and \(B\), the non-separability means the impossibility of factorizing the density matrix of the combined system \( \rho \) as \( \rho = \sum_{k} \lambda_k \rho^k_A \rho^k_B  \), with \( \sum_{k} \lambda_k = 1\). In \cite{duan2000inseparability}, a criterion for inseparability was developed by Duan et al., which provides a sufficient condition for the entanglement of any two party continuous variable states \cite{simon2000peres}. {For two radiation fields with annihilation operators $a$ and $b$, this criterion translates to
  \begin{equation}
  \mathcal{D}_{ab} = 4(\Delta X_{ab})^2 + 4(\Delta Y_{ab})^2 - 2 < 0,\label{eq:duan}
  \end{equation}
  where $(\Delta X_{ab})^2$ and $(\Delta Y_{ab})^2$ are defined in Eq. (\ref{IMS}). The presence of squeezing  does not ensure the existence of entanglement as at a given time squeezing can happen only in one quadrature. Thus, this criterion captures the asymmetry in the fluctuations in $X$ and $Y$ and this is why it's studied independently.}
  In what follows, we refer to this criterion of entanglement as Duan's criterion.
 
 \item \textit{Hillery-Zubairy(HZ) criteria of entanglement}: In \cite{hillery2006entanglement}, it was shown that for two field modes \(a\) and \(b\), two inseparability criteria are
 \begin{equation}
 \mathcal{E}_{ab} = \langle a^\dagger a b^\dagger b \rangle - |\langle a b^\dagger  \rangle|^2 < 0,  \label{E}
 \end{equation} 
and
 \begin{equation}
\tilde{ \mathcal{E}}_{ab} = \langle a^\dagger a \rangle \langle b^\dagger b \rangle - |\langle a b  \rangle |^2< 0.\label{Etilde}
 \end{equation}
 
 \item \(Steering\): The notion of steering, as an apparent action at a distance, was introduced by Schr\"{o}dinger while discussing the $EPR$ paradox \cite{schrodinger1935discussion}, and shares logical differences both with non-separability and Bell non-locality. While as non-separability and Bell non-locality are symmetric between two parties, say Alice and Bob, steering is inherently asymmetric, addressing whether Alice can change the state of Bob's system by applying local measurements. An operational definition of steering was first provided in \cite{wiseman2007steering}, wherein they proved that steerable states are a strict subset of the entangled states and a strict superset of the states that can exhibit Bell non-locality.  In the context of field modes $a$ and $b$, the $EPR-steering~entanglement$ is confirmed if it satisfies \cite{he2012einstein}
 
 \begin{equation}
 0 < 1 + \frac{\langle a^{\dagger } a b^{\dagger } b\rangle - |\langle a b^{\dagger } \rangle|^2}{\langle a^{\dagger }a(b b^{\dagger } - b^{\dagger } b) \rangle} <\frac{1}{2}. \label{eq:St1}
 \end{equation}
This result can be proved by the methods given in  \cite{cavalcanti2011unified}. The above steering condition (\ref{eq:St1}) can be expressed in terms of the HZ criterion Eq.~(\ref{E}), the condition reads:
  \begin{equation} \label{Steering}
  \mathcal{S}_{AB} = \mathcal{E}_{ab} +  \frac{\langle a^{\dagger} a \rangle }{2} < 0.
 \end{equation}
 The concept of steering being inherently asymmetric \cite{cavalcanti2009experimental}, it will be interesting to compare $\mathcal{S}_{AB}$ and $\mathcal{S}_{BA} =  \mathcal{E}_{ab} +  \frac{\langle b^{\dagger} b \rangle }{2} $.

  \item \textit{Multimode entanglement}: In \cite{li2007entanglement}, a class of inequalities was derived for detecting the entanglement in multimode systems. In the case of a tripartite state, viz., the one corresponding to the three  modes $a$, $b,$ and $c$, the sufficient conditions for not being bi-separable of the form $ab|c$ (in which a compound mode $ab$ is entangled with mode $c$), are given as follows:
  \begin{align}
   E_{ab|c} &=  \langle a^{\dagger} a b^{\dagger} b c^{\dagger} c \rangle - |\langle a b c^{\dagger} \rangle|^2<0 \label{Bisep1},\\
  E_{ab|c}^{\prime} &=   \langle a^{\dagger} a b^{\dagger} b\rangle\langle c^{\dagger} c \rangle - |\langle a b c \rangle|^2 <0      \label{Bisep2}.
  \end{align}
A three-mode quantum state is fully entangled by the satisfaction of either or both of the following sets of inequalities:
  \begin{equation}
E_{ab|c} < 0, \quad E_{bc|a} < 0, \quad  E_{ac|b} < 0, \label{set1}
  \end{equation}
   \begin{equation}
  E_{ab|c}^{\prime} < 0, \quad E_{bc|a}^{\prime} < 0, \quad  E_{ac|b}^{\prime} < 0. \label{set2}
  \end{equation}
\end{itemize}
 It is worth mentioning here that the analysis of the above mentioned witnesses of non-classicality involves higher order products of the operators. These higher order correlations can be decorrelated by the prescription given in  \cite{anglin2001dynamics}. Specifically, in what follows, we have made use of  $\langle \hat{a} \hat{b} \hat{c} \rangle \approx \langle \hat{a} \hat{b} \rangle \langle \hat{c} \rangle + \langle \hat{a} \rangle \langle \hat{b} \hat{c} \rangle + \langle \hat{a} \hat{c} \rangle \langle \hat{b} \rangle - 2 \langle \hat{a} \rangle \langle \hat{b} \rangle \langle\hat{c} \rangle$, which basically makes use of the Bogoliubov theory of linearized quantum corrections to mean field effects.

\section{Results and Discussion} \label{Results}

In this section, we study the non-classical properties of our system as manifested through various witnesses of non-classicality discussed above. The analysis is carried out by placing the ensembles in the four configurations, viz., AA, AN, NA, and NN configurations.  However, the analysis performed for NN and AA modes are not as detailed as in AN and NA configurations. The investigation performed for all the configurations is summarized in Table \ref{tab1}, for the convenience of the reader. It clearly emerges that the AA configuration is more suited for observing the various facets of non-classicality in the system. {In some cases, other configurations may be preferred due to sufficient depth of the nonclassciality witness, which is desired in some particular applications having practical relevance.} The effect of external driving field on the various non-classical witnesses is studied with respect to $\Delta t$, where $\Delta$  is the common detuning for the three modes $\hat{A}$, $\hat{B}$ and $\hat{C}$.  The various parameters used for AN configuration  are $G_{A} = 0.2\Delta$, $G_{B} = 0.02\Delta$, $\Gamma_{A} = 2\Delta$, $\Gamma_{B} = 0.2\Delta$, and $\Gamma_{C} = 0.2\Delta$. For NA configuration, $G_{A} = 0.02\Delta$, $G_{B} = 0.2\Delta$, $\Gamma_{A} = 0.2\Delta$, $\Gamma_{B} = 2\Delta$, and $\Gamma_{C} = 0.2\Delta$. For AA configuration, $G_{A} = 0.2\Delta$, $G_{B} = 0.2\Delta$, $\Gamma_{A} = 2\Delta$, $\Gamma_{B} = 2\Delta$, and $\Gamma_{C} = 0.2\Delta$. And finally, for NN configuration $G_{A} = 0.02\Delta$, $G_{B} = 0.02\Delta$, $\Gamma_{A} = 0.2\Delta$, $\Gamma_{B} = 0.2\Delta$, and $\Gamma_{C} = 0.2\Delta$. 
In all the cases, we have considered here vacuum bath.

The initial conditions (at $t=0$) are chosen in such a way  that the average number of photons in the cavity field and the average number of excitations in the two ensembles are all equal to 1. Figure \ref{number} shows the evolution of the average number of bosons corresponding to the driven ensemble mode ($ \langle \hat{A}^\dagger\hat{A} \rangle$), the undriven ensemble mode ($ \langle \hat{B}^\dagger \hat{B} \rangle$) and the average number of the cavity photons ($ \langle \hat{C}^\dagger \hat{C} \rangle$).  The average number of excitations is found to  drop quickly for the ensemble placed at the Antinode of the cavity field, compared to the ensemble placed at the Node of the cavity field. One can also see vivid variations in the average excitation number of the driven ensemble, when placed at the Node of the cavity field. In other words, placing the ensemble at the Antinode of  the cavity field, shadows the effect of the external field. We have not shown similar variation in the boson number for remaining two configurations as it is quite similar to what is observed here. The interested readers are referred to the supplemental material \cite{supp} for various results summarized in Table \ref{tab1}, but not illustrated in the main paper.

\begin{widetext}

	\begin{table}[h!]
		\centering
		\caption{\label{tab1} Various witnesses of non-classicality that are investigated in this paper for different configurations, both in the absence as well as in the presence of external field characterized by $\chi$. Here, a tick indicates the presence of a non-classical feature characterized by the non-classicality witness mentioned in the first column of the same row, while a cross is the indicator of failure in the detection of that non-classicality feature. It clearly emerges that the AA configuration is more suited for observing various facets of non-classicality in the system. {However, other configurations can be preferred in some cases as far as the depth of nonclassciality witnesses is concerned.}}
		\label{summary}
		\begin{tabular}{|p{3.9cm}|p{1.4cm}p{1.4cm}|p{1.4cm}p{1.4cm}|p{1.4cm}p{1.4cm}|p{1.4cm}p{1.4cm}|}
			\hline
			\multirow{2}{*}{Witnesses of non-classicality}                      &              \multicolumn{2}{c|}{AN configuration}                                                                                                              & \multicolumn{2}{c|}{NA configuration}                                                                       & \multicolumn{2}{c|}{AA configuration}                                                                                        & \multicolumn{2}{c|}{NN configuration}                                 \\ 
			& $\chi=0$                                           & $\chi=0.2\Delta$                                                                              & $\chi=0$                                              & $\chi=0.2\Delta$                                                      & $\chi=0$                                     & $\chi=0.2\Delta$                                                  & $\chi=0$                                                          & $\chi=0.2\Delta$                  \\ \hline \hline
			Mandel parameter                                   & ($\AT$, $\BT$, $\CT$)                                                  & ($\AT$, $\BT$, $\CT$)                                                & ($\AT$, $\BT$, $\CT$)                                                    & ($\AT$, $\BT$, $\CT$)                               & ($\AT$, $\BT$, $\CT$)                             & ($\AT$, $\BT$, $\CT$)                                & ($\AT$, $\BT$, $\CT$)                                                     & ($\AT$,$\BT$,$\CT$)           \\ \hline                               
			Singlemode Squeezing                               & ($\AT$, $\BT$, $\CT$)                                                  & ($\AT$, $\BT$, $\CT$)                                                & ($\AT$, $\BT$, $\AT$)                                                    & ($\AT$, $\BT$, $\AT$)                               & ($\AT$, $\BT$, $\AT$)                             & ($\AT$, $\BT$, $\AT$)                                & ($\AT$, $\BT$, $\AT$)                                                     & ($\AT$, $\BT$, $\AT$)         \\ \hline
			Intermodal Squeezing                               & ($\ABT$, $\BCT$, $\ACT$)                                               & ($\ABT$, $\BCT$, $\ACT$)                                             & ($\ABT$, $\BCT$, $\ACT$)                                                 & ($\ABT$, $\BCT$, $\ACT$)                            & ($\ABT$, $\BCT$, $\ACT$)                          & ($\ABT$, $\BCT$, $\ACT$)                             & ($\ABT$, $\BCT$, $\ACT$)                                                  & ($\ABT$, $\BCT$, $\ACT$)      \\ \hline
			Hillery Zubairy criterion ($\mathcal{E}$)          & ($\ABT$, $\BCT$, $\ACT$)                                               & ($\ABT$, $\BCT$, $\ACT$)                                             & ($\ABT$, $\BCT$, $\ACT$)                                                 & ($\ABT$, $\BCT$, $\ACT$)                            & ($\ABT$, $\BCT$, $\ACT$)                          & ($\ABT$, $\BCT$, $\ACT$)                             & ($\ABT$, $\BCT$, $\ACT$)                                                  & ($\ABT$, $\BCT$, $\ACT$)      \\ \hline
			Hillery Zubairy criterion ($\tilde{\mathcal{E}}$)  & ($\ABT$, $\BCT$, $\ACT$)                                               & ($\ABT$, $\BCT$, $\ACT$)                                             & ($\ABT$, $\BCT$, $\ACT$)                                                 & ($\ABT$, $\BCT$, $\ACT$)                            & ($\ABT$, $\BCT$, $\ACT$)                          & ($\ABT$, $\BCT$, $\ACT$)                             & ($\ABT$, $\BCT$, $\ACT$)                                                  & ($\ABT$, $\BCT$, $\ACT$)      \\ \hline
			Duan criterion                                     & ($\ABT$, $\BCT$, $\ACT$)                                               & ($\ABT$, $\BCT$, $\ACT$)                                             & ($\ABT$, $\BCT$, $\ACT$)                                                 & ($\ABT$, $\BCT$, $\ACT$)                            & ($\ABT$, $\BCT$, $\ACT$)                          & ($\ABT$, $\BCT$, $\ACT$)                             & ($\ABT$, $\BCT$, $\ACT$)                                                  & ($\ABT$, $\BCT$, $\ACT$)      \\ \hline
			Biseperability criterion (E)                        & ($\ABcT$, $\BCaT$, $\ACbT$)                                            & ($\ABcT$, $\BCaT$, $\ACbT$)                                          & ($\ABcF$, $\BCaF$, $\ACbF$)\cellcolor[gray]{0.8}                         & ($\ABcF$, $\BCaF$, $\ACbF$)\cellcolor[gray]{0.8}    & ($\ABcT$, $\BCaT$, $\ACbT$)                       & ($\ABcT$, $\BCaT$, $\ACbT$)                          & ($\ABcT$, $\BCaT$, $\ACbT$)                                               & ($\ABcT$, $\BCaT$, $\ACbT$)   \\ \hline
			Biseperability criterion ($E^\prime$)               & ($\ABcT$, $\BCaT$, $\ACbT$)                                            & ($\ABcT$, $\BCaT$, $\ACbT$)                                          & ($\ABcT$, $\BCaT$, $\ACbT$)                                              & ($\ABcT$, $\BCaT$, $\ACbT$)                         & ($\ABcT$, $\BCaT$, $\ACbT$)                       & ($\ABcT$, $\BCaT$, $\ACbT$)                          & ($\ABcT$, $\BCaT$, $\ACbT$)                                               & ($\ABcT$, $\BCaT$, $\ACbT$)   \\ \hline
			Single mode antibunching                           & ($\AT$, $\BT$, $\CT$)                                                  & ($\AT$, $\BT$, $\CT$)                                                & ($\AT$, $\BT$, $\CT$)                                                    & ($\AT$, $\BT$, $\CT$)                               & ($\AT$, $\BT$, $\CT$)                             & ($\AT$, $\BT$, $\CT$)                                & ($\AT$, $\BT$, $\CT$)                                                     & ($\AT$,$\BT$,$\CT$)           \\ \hline
			Intermodal antibunching                            & ($\ABT$, $\BCT$, $\ACT$)                                               &($\ABT$, $\BCT$, $\ACT$)                                              & ($\ABT$, $\BCT$, $\ACF$)\cellcolor[gray]{0.8}                            & ($\ABT$, $\BCT$, $\ACT$)                            & ($\ABT$, $\BCT$, $\ACT$)                          & ($\ABT$, $\BCT$, $\ACT$)                             & ($\ABT$, $\BCT$, $\ACF$)\cellcolor[gray]{0.8}                             & ($\ABT$, $\BCT$, $\ACT$)      \\ \hline
			Steering                                           & ($\ABT$,$\BAT$, $\BCT$,$\CBT$, $\ACF$, $\CAF$)\cellcolor[gray]{0.8}    & ($\ABT$,$\BAT$, $\BCT$,$\CBT$, $\ACF$, $\CAF$)\cellcolor[gray]{0.8}  & ($\ABF$,$\BAT$, $\BCT$,$\CBT$, $\ACF$, $\CAF$)\cellcolor[gray]{0.8}      & ($\ABF$,$\BAT$, $\BCT$,$\CBT$, $\ACT$, $\CAT$)      & ($\ABT$,$\BAT$, $\BCT$,$\CBT$, $\ACT$, $\CAT$)    & ($\ABT$,$\BAT$, $\BCT$,$\CBT$, $\ACT$, $\CAT$)       & ($\ABF$,$\BAF$, $\BCT$,$\CBT$, $\ACF$, $\CAF$)\cellcolor[gray]{0.8}       & ($\ABT$,$\BAT$, $\BCT$,$\CBT$, $\ACT$, $\CAF$)\cellcolor[gray]{0.8}  \\ \hline                           	
		\end{tabular}
	\end{table}

\end{widetext}

\FloatBarrier

\begin{figure}[ht] 
	\centering
\includegraphics[width=70mm]{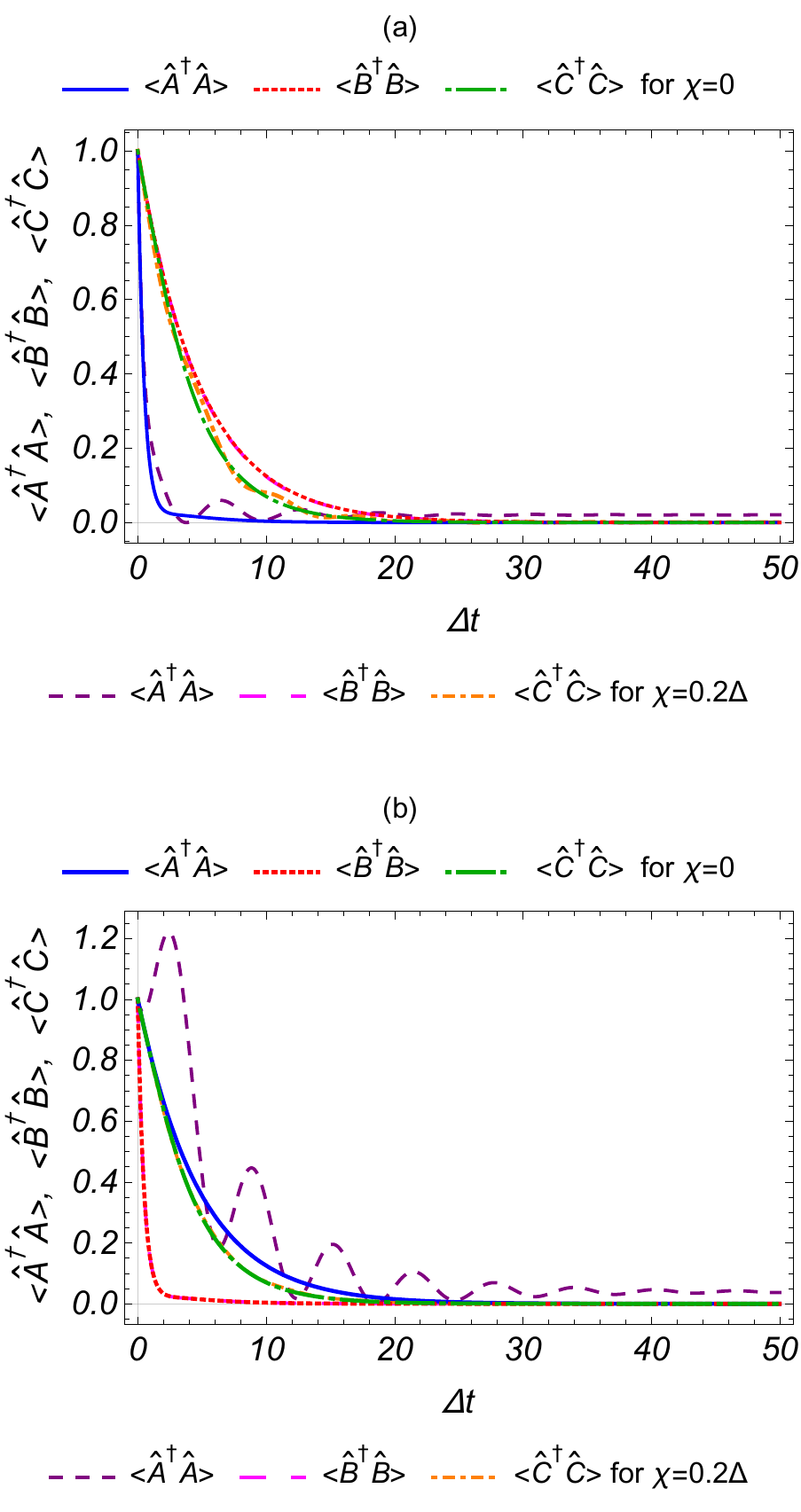}
	\caption{(Color online) Average number of cavity photons and excitations corresponding to the two ensembles, studied with respect to the dimensionless parameter $\Delta t$. (a) and (b)  correspond to AN (left ensemble at Antinode and the right ensemble at Node) and NA (left ensemble at Node and the right ensemble at Antinode) configurations, respectively. The average number of excitations corresponding to the driven ensemble ($\langle A^{\dagger} A \rangle$), the undriven ensemble ($\langle B^{\dagger} B \rangle$) and the average cavity photon number ($\langle C^{\dagger} C \rangle$) is depicted for $\chi = 0$ and $\chi = 0.2\Delta$. {All the quantities shown in the plots in the present paper are dimensionless.}}
	\label{number}
\end{figure}

 Evolution of the average boson number discussed above gives us a feeling of the system dynamics, but does not provide us any information about the non-classical nature of the system. To obtain the non-classical characteristics of the system, we begin with the study of variation of a single mode non-classicality witness known as Mandel parameter $Q_M$, which has been   introduced in the previous section.  Variation of $Q_M$ with respect to rescaled time $\Delta t$ is plotted in Fig. (\ref{mandel}) for all three modes of the system in AN and NA configurations. The condition for non-classicality  is implied by the negative values of $Q_M$ (i.e., $Q_M < 0$) and can be interpreted as the sub-Poissonian statistics of the corresponding field.  Further, it is observed that the application of the external field to the driven ensemble makes $Q_M $ more negative, and thus the driving optical field may be used to enhance the amplitude of the non-classiclaity witness in both the driven ensemble and cavity mode. Negative values of $Q_M $ are also observed for AA and NN configurations and similar inferences can be drawn from it as mentioned in Table \ref{tab1}; the illustrations of the  results are  given in the supplementary material.
 Specifically, we found  non-classicality in the driven (undriven) ensemble in the absence of the external drive. Although in the absence of the driving term, the Hamiltonian given by Eq. (\ref{H1}) is symmetric for both the ensembles, the observed behavior can be attributed to different values of decay constants for the modes under consideration. Note that the non-classicality observed in the driven ensemble for higher intensity of the driving field establishes that driving field can be used to control the amount of non-classicality in the system. 
	\begin{figure}[ht] 
	\centering
    \includegraphics[width=70mm]{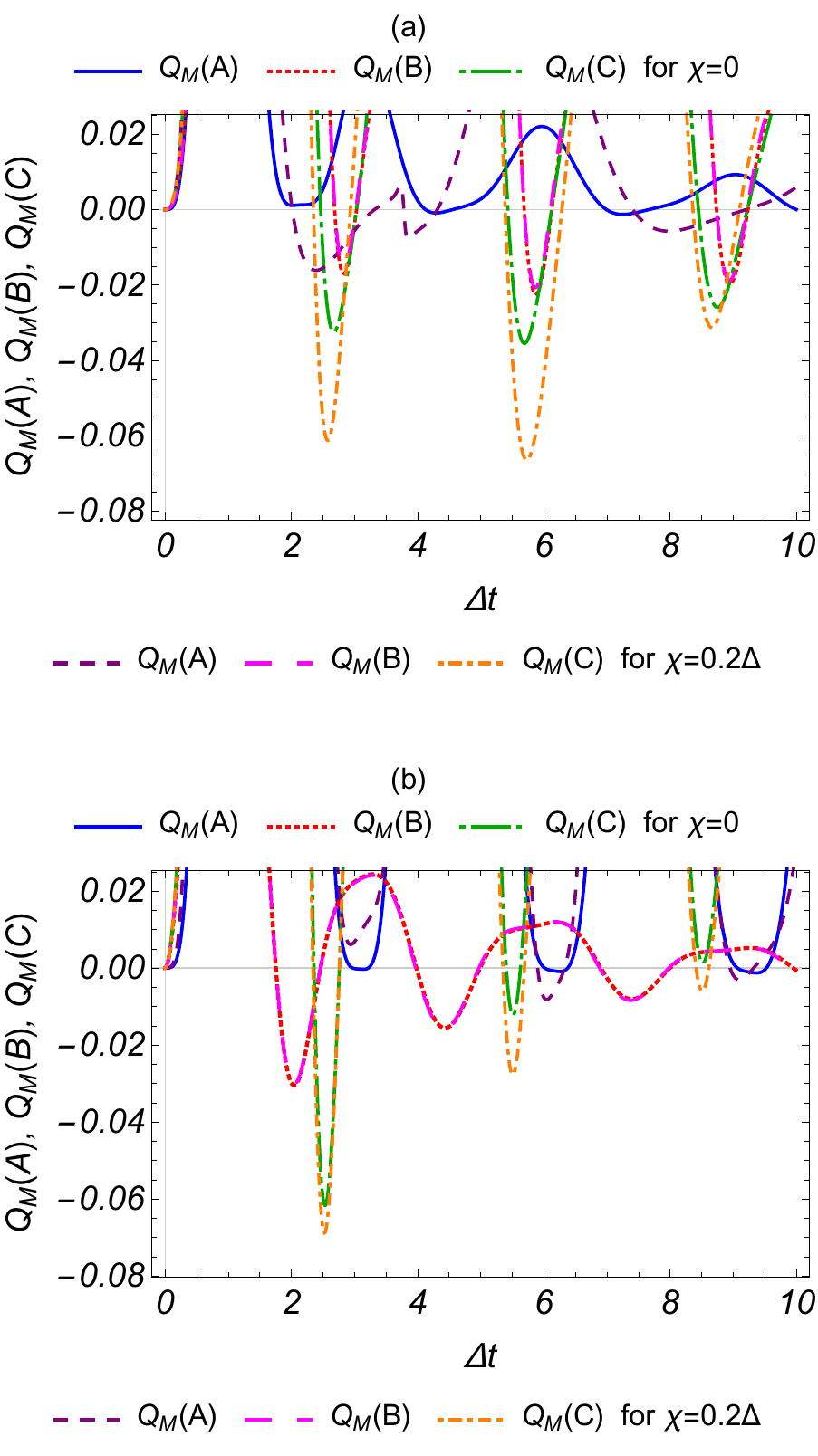}
	\caption{(Color online) Mandel parameter with respect to the dimensionless parameter $\Delta t$. Figures (a) and (b) correspond to AN and NA configurations, respectively. The non-classical nature of the field corresponding to the mode $\alpha$  is confirmed by $Q_M(\alpha) < 0$.}
	\label{mandel}
\end{figure}

Motivated by the presence of single mode non-classicality in the boson number distribution also illustrating the presence of single mode antibunching, we also study the possibilities of compound mode antibunching using Eq. (\ref{InterModalAntiBun}). Fig. (\ref{IMAB}) shows the variation  of non-classicality parameter for the intermodal antibunching as defined by Eq. (\ref{InterModalAntiBun}) for all possible compound modes in AN and NA configuration. The criterion $\mathcal{A}_{\alpha \beta} < 0$ is satisfied for all the modes $\alpha/ \beta \in \{\hat{A}, \hat{B}, \hat{C}\}$. One can see the enhancement in the depth of intermodal antibunching parameter by the action of the external field driving the ensemble $A$. Similar studies in the case of AA and NN configurations also show intermodal antibunching as summarized in Table \ref{tab1}.

		\begin{figure}[ht] 
		\centering
	    \includegraphics[width=70mm]{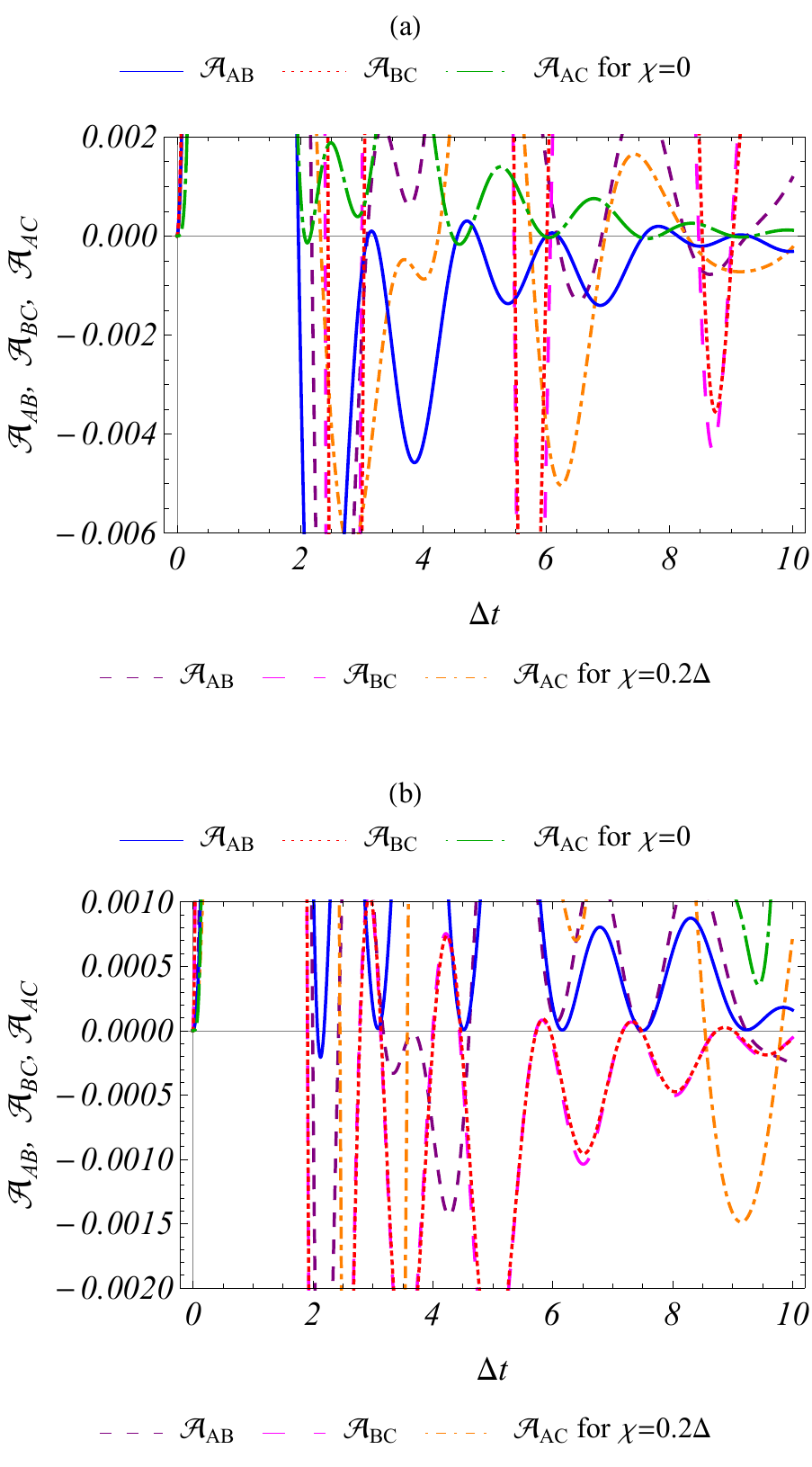}
		\caption{(Color online) Showing intermodal antibunching parameter $\mathcal{A}_{\alpha \beta}$ for modes $\alpha$ and $\beta$, plotted with respect to the parameter  $\Delta t$. (a) and (b) correspond to AN and NA configurations, respectively. The existence of intermodal antibunching is confirmed if $\mathcal{A}_{\alpha \beta} < 0$. }
		\label{IMAB}
	\end{figure}

\begin{widetext}

\begin{figure}[t]
	\centering
	\includegraphics[width=180mm]{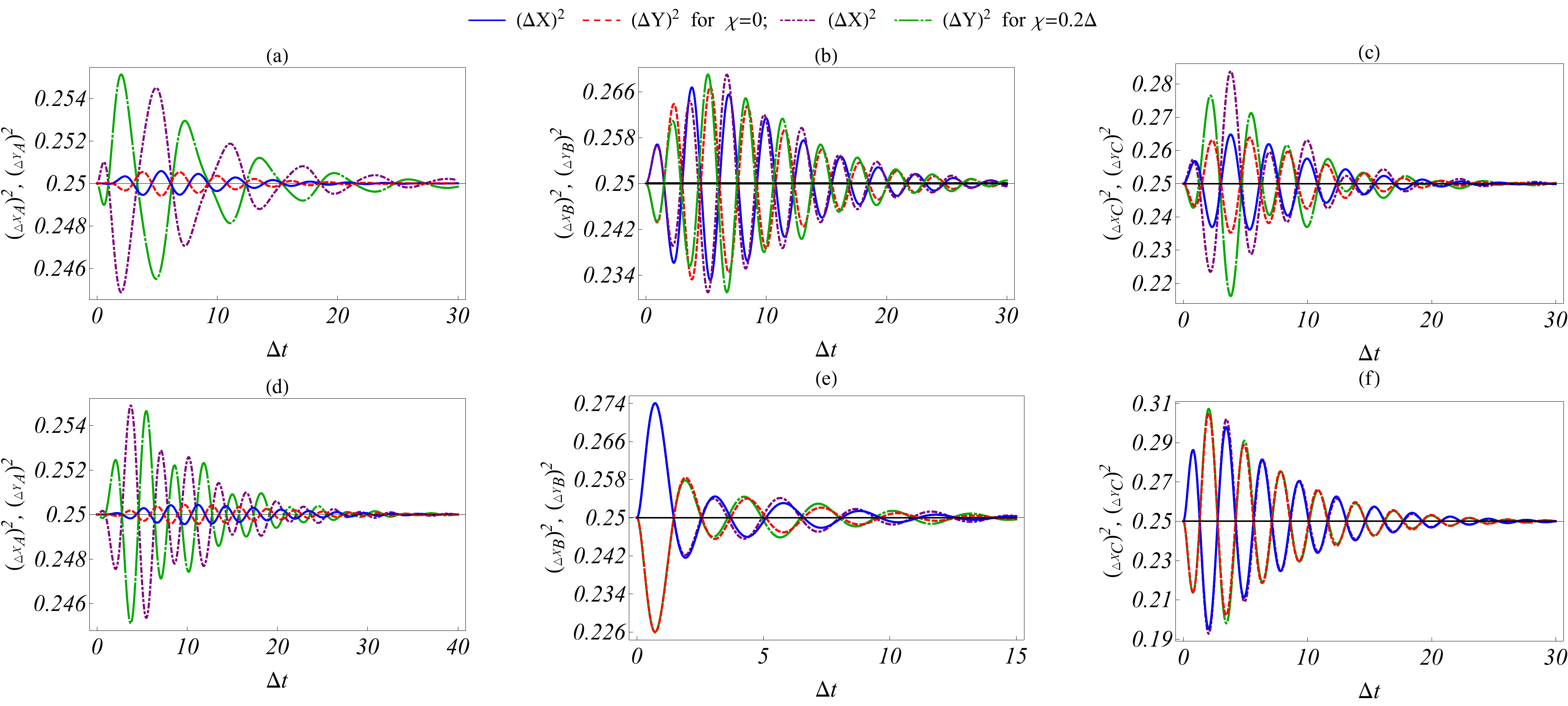}
	\caption{(Color online) Single mode squeezing, as defined in Eqs. (\ref{SMSX}) and (\ref{SMSY}), plotted with respect to the dimensionless parameter $\Delta t$. Sub-figures (a)-(d), (b)-(e) and (c)-(f) correspond to modes $\hat{A}$, $\hat{B}$ and $\hat{C}$, respectively. The top and bottom panels pertain to the AN and NA configurations, respectively. It is clear that the application of the external field to the driven ensemble ($\hat{A}$), that is, the non-zero value of $\chi$, enhances the squeezing in the respective quadratures of a particular mode.}
	\label{SingleModeSqueezing}
\end{figure}

\end{widetext}

\begin{figure}[ht] 
	\centering
	\includegraphics[width=80mm]{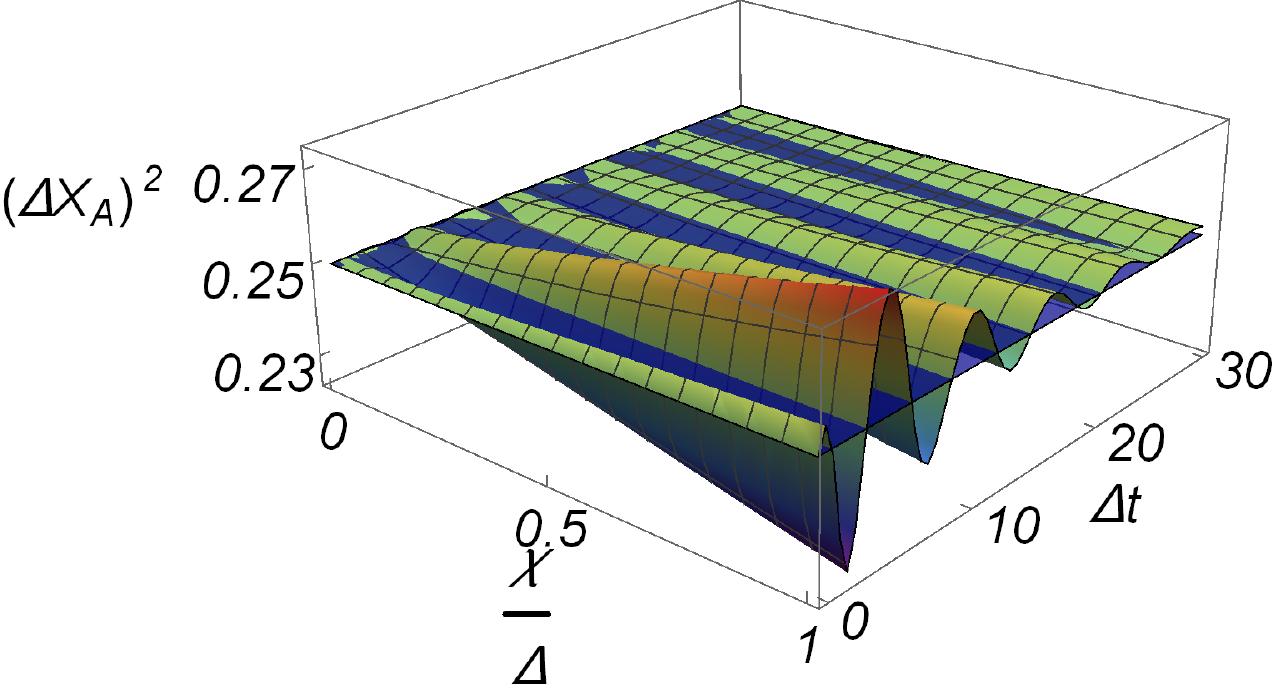}	
	\caption{(Color online). Showing squeezing parameter $(\Delta X_A)^2$ for mode $\hat{A}$ as a function of the driving field strength $\chi$ as well as the dimensionless parameter $\Delta t$. The enhancing effect in quadrature squeezing as a result of increase in the strength of the driving field is observed.}
	\label{3DsqueezXA}
     \end{figure}

 After witnessing the signatures of non-classicality in  all three modes of the system through the negative values of the Mandel $Q_M$ parameter, we turn our attention towards single mode squeezing; criterion for which is defined in Eq. (\ref{SMSX}) and Eq. (\ref{SMSY}). Fig. \ref{SingleModeSqueezing} illustrates the presence of  the quadrature squeezing in all the individual modes, both in AN and NA configurations. A similar study for AA and NN configurations is carried out and the results (not displayed here) are summarized in Table \ref{tab1}. The field mode $\hat{A}$, corresponding to the driven ensemble, shows an appreciable enhancement in the magnitude of squeezing illustrated by the decrease of the variance in one quadrature with respect to the coherent state level as soon as the external field is applied. This enhancement is also observed in the undriven mode $\hat{B}$ and the cavity mode $\hat{C}$, but with relatively lesser magnitude in AN configuration, while as in NA configuration, the enhancement in the non-classicality of field modes $\hat{B}$ and $\hat{C}$ is quite meager. This can be attributed to the fact that in NA configuration, the driven ensemble, being at the node of the cavity field, is weakly coupled  to it. Therefore, we  conclude that the amount of non-classicality in the driven ensemble can be controlled by the strength of the driving field, however, the driven ensemble should be placed at the Antinode of the cavity field (AN or AA configurations) for this control to be effective on the cavity field mode ($\hat{C}$) and the undriven ensemble excitation mode ($\hat{B}$), as well. { However, in some cases (as in Fig. \ref{SingleModeSqueezing} (b)), with an increase in the strength of the external field nonclassicality present in the absence of external field decreases initially for a small period of time before increasing thereafter. This behavior could not be explained by the present study and may be attempted in the near future. The role of external driving field strength as a control parameter can be further established using Fig. \ref{3DsqueezXA}, which illustrates the variation of $(\Delta X_a)^2$ with respect to the external driving field strength and time. The enhancing effect of the external field on the quadrature squeezing is clearly visible in this case.}

\begin{figure}[t]
	\centering
	\includegraphics[width=70mm]{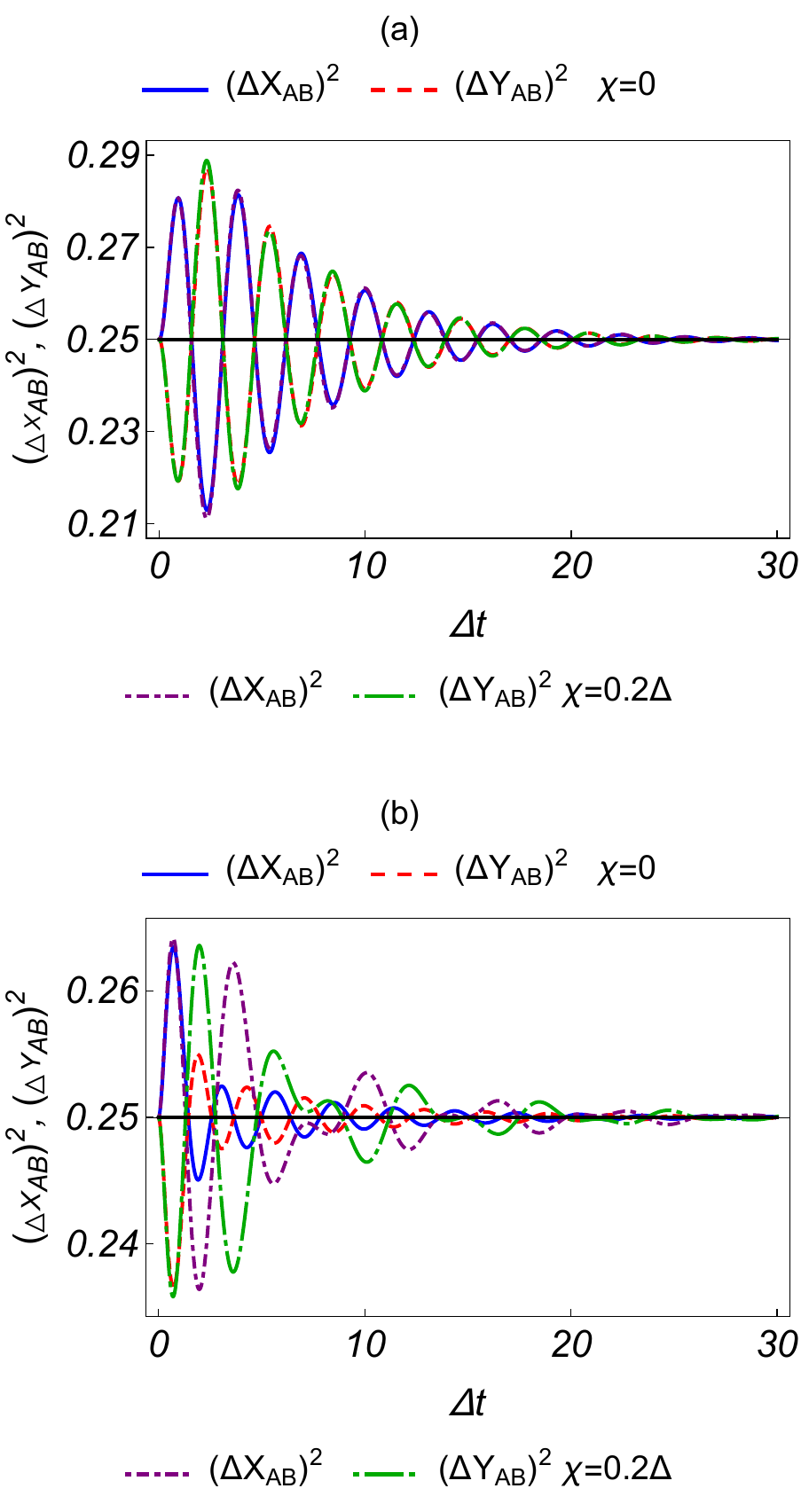}
	\caption{(Color online) Showing inter-modal squeezing, defined by Eq. (\ref{IMS1}), with respect to $\Delta t$. (a) and (b) show the squeezing in compound mode $\hat{A}\hat{B}$, in AN and NA configurations, respectively. One finds enhancement in the intermodal squeezing as a result of driving ensemble ($\hat{A}$) by the application of the external field.}
	\label{InterModalSqueezing}
\end{figure}

   Motivated by the observation for the single mode squeezing, we investigated the presence of intermodal squeezing using the criterion given in Eqs. (\ref{IMS1}) and (\ref{IMS2}). The outcome of the investigation is plotted in  Fig. \ref{InterModalSqueezing}, which clearly shows  the existence of intermodal squeezing in the compound mode $\hat{A}\hat{B}$. One can observe the  amplification in the  squeezing parameters as a consequence of an increase in the external field driving the atomic ensemble $\mathcal{S}_{A}$ (cf. Fig. \ref{InterModalSqueezing} (b)). A similar study for NN and AA configuration is also carried out with compound modes $\hat{B}\hat{C}$ and  $\hat{A}\hat{C}$ in all four configurations. The presence of squeezing in different possible compound ensemble-ensemble and ensemble-cavity modes in all the four configurations is observed and is summarized in Table \ref{tab1}. It is worth mentioning here that the enhancement in the values of the witness of the intermodal squeezing is found to be more prominent in the  compound mode  $\hat{A}\hat{C}$ when compared with $\hat{A}\hat{B}$ or $\hat{B}\hat{C}$. This can be attributed to the fact that the amount of non-classicality in mode $\hat{B}$ is less susceptible to the driving field.

Nonclassical features manifested through the negative values of Mandel $Q_M$ parameter, intermodal antibunching and the criteria of single mode and compound mode squeezing have been studied since long using various techniques including short-time approach \cite{perina1995photon,perina1991quantum} and Sen-Mandal approach \cite{thapliyal2014higher,thapliyal2017nonclassicality}, but most of those studies were limited to closed system configuration. In the present work, we have reported the existence and dynamics of these non-classical features in the backdrop of open quantum systems. To continue the investigation further, we may note that among various non-classical features, entanglement has drawn maximum attention of the scientific community because of its enormous applications in quantum computing and communication and because of the fact that it may lead to various phenomenon having no classical analogue such as, dense coding \cite{bennett1992communication} and teleportation \cite{bennett1993teleporting}. Keeping this in mind, we would now look into the possibility of observing intermodal entanglement in the system of our interest. To do so, we will use a set of inseparability criteria, each of which is only sufficient and consequently when one of the criteria fails, another one may succeed to detect entanglement. We begin with Duan's criterion for inseparability defined in Eq. (\ref{eq:duan}) and graphically present the obtained results in Fig. \ref{duan}. { It is clear from Fig. \ref{duan}  that  the condition for inseparability (i.e., $\mathcal{D}_{\alpha \beta} < 0$) is satisfied for  modes $\alpha$ and  $\beta$. Irrespective of whether the driven ensemble is placed at Node or Antinode of the cavity field, the value of the Duan parameters $\mathcal{D}_{AB}$ and $\mathcal{D}_{BC}$ is very small. In the NA configuration (shown in Fig. \ref{duan} (b)), the Duan parameter $\mathcal{D}_{A B}$, $\mathcal{D}_{B C}$ as well as $\mathcal{D}_{A C}$ become negative, thereby witnessing the presence of entanglement among all the modes. On the other hand, in AN configuration, $\mathcal{D}_{BC}$ is non-negative (cf. Fig. \ref{duan} (a)). This implies that transforming the system from AN to NA configuration enhances the  intermodal entanglement, which can also be viewed in the enhancement of the Duan parameter $\mathcal{D}_{A C}$ in NA configuration.}

\begin{figure}[ht] 
	\centering
    \includegraphics[width=70mm]{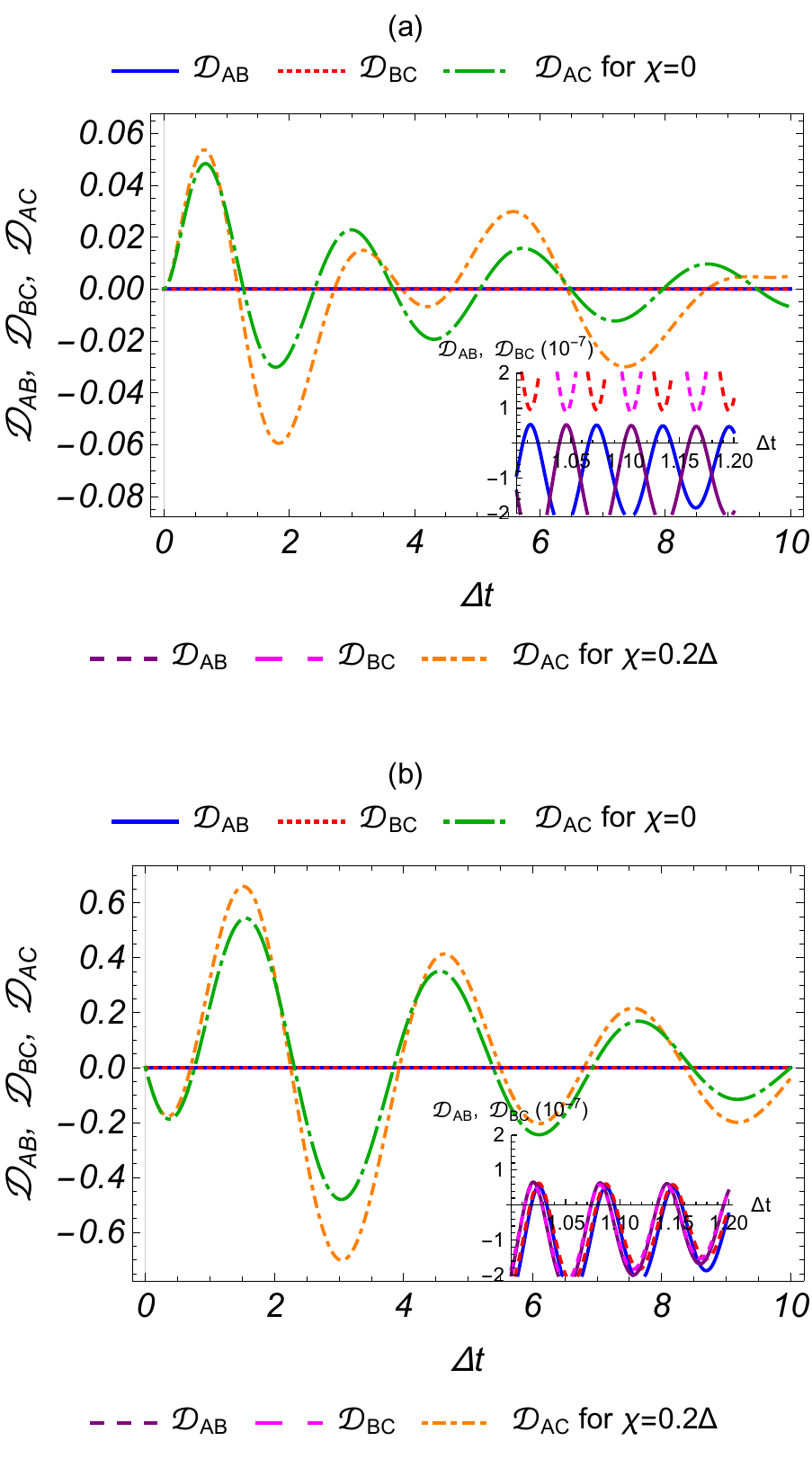}
	\caption{(Color online) Showing Duan separability parameter against $\Delta t$. Figures (a) and (b) correspond to AN and NA configurations, respectively. The sufficient condition for inseparability is implied by $\mathcal{D}_{\alpha \beta} < 0$, for modes $\alpha$ and $\beta$.}
	\label{duan}
\end{figure}

\begin{figure*}[ht] 
	\centering
	\includegraphics[width=140mm]{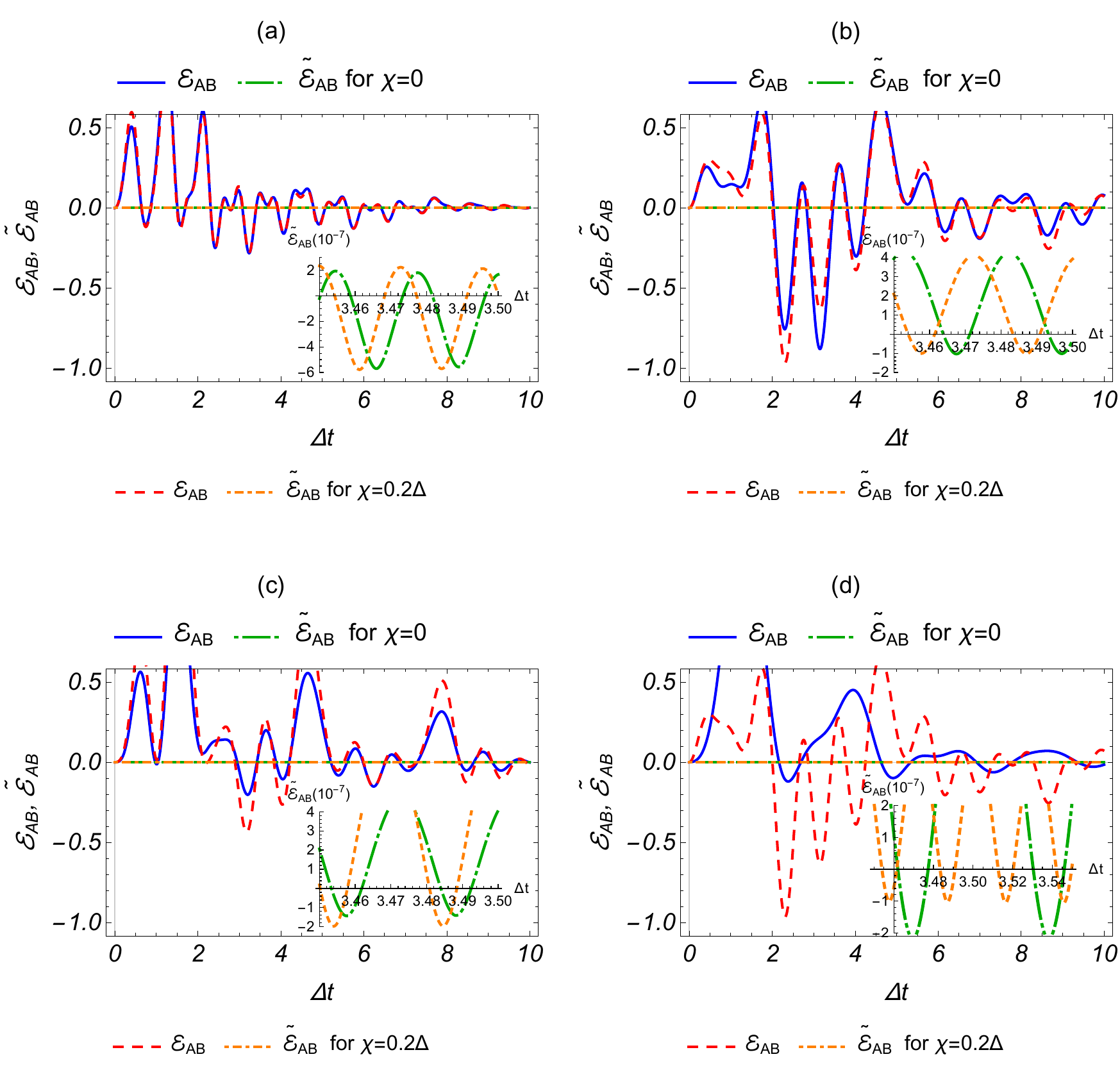}
	\caption{(Color online) Hillery Zubairy  criteria, as defined in Eqs. (\ref{E}) and (\ref{Etilde}), plotted against the dimensionless parameter $\Delta t$ for modes $\hat{A}$ and $\hat{B}$. The figures (a), (b), (c) and (d) correspond to the AA, AN, NA and NN  configurations, respectively. The negative values of the Hillery Zubairy parameters ($\rm HZPs$), viz., $\mathcal{E}_{AB}$ and $\tilde{\mathcal{E}}_{AB}$ provide the sufficient condition for the entanglement between the corresponding modes. The nonzero value of $\chi$ makes $\rm HZPs$ more negative at certain points, and hence, enhance the entanglement between the corresponding modes.}
	\label{HZ}
\end{figure*}

    As stated above, moment-based criteria of inseparability are only sufficient. Hence, it makes sense to look into the possibility of observing entanglement in light of one or more different criteria. We study the famous  Hillery-Zubariy criteria  defined by Eqs. (\ref{E}) and (\ref{Etilde}) and have illustrated the corresponding results in Fig. \ref{HZ}, where negative values of $\mathcal{E}_{AB}$ and $\tilde{\mathcal{E}}_{AB}$ confirm the existence of the intermodal entanglement between modes $\hat{A}$ and $\hat{B}$  for all configurations. At times, weak signatures of entanglement are found through $\tilde{\mathcal{E}}_{AB}$ criterion, but relatively stronger signatures are found through $\mathcal{E}_{AB}$ criterion (see Fig. (\ref{HZ})). Similarly, one may observe in Fig. (\ref{duan})-(a) and (\ref{HZ})-(b) that, at $\Delta t=6$, Duan's criterion failed to detect entanglement but is captured by the Hillery-Zubairy criteria.  Further, the study revealed the relevance of placement of ensembles in the cavity for the generation  of entanglement between two spatially separated ensembles interacting with a common  cavity field. Also, going from AA to NN configuration, the effect of external driving field becomes relevant for controlling the amount of entanglement. A similar study for the remaining two compound modes 
 $\hat{B}\hat{C}$ and  $\hat{A}\hat{C}$ also established that they are always entangled in all configurations as summarized in Table \ref{tab1} (see the supplementary material for detail).

\begin{figure*}[ht] 
	\includegraphics[width=0.85\linewidth]{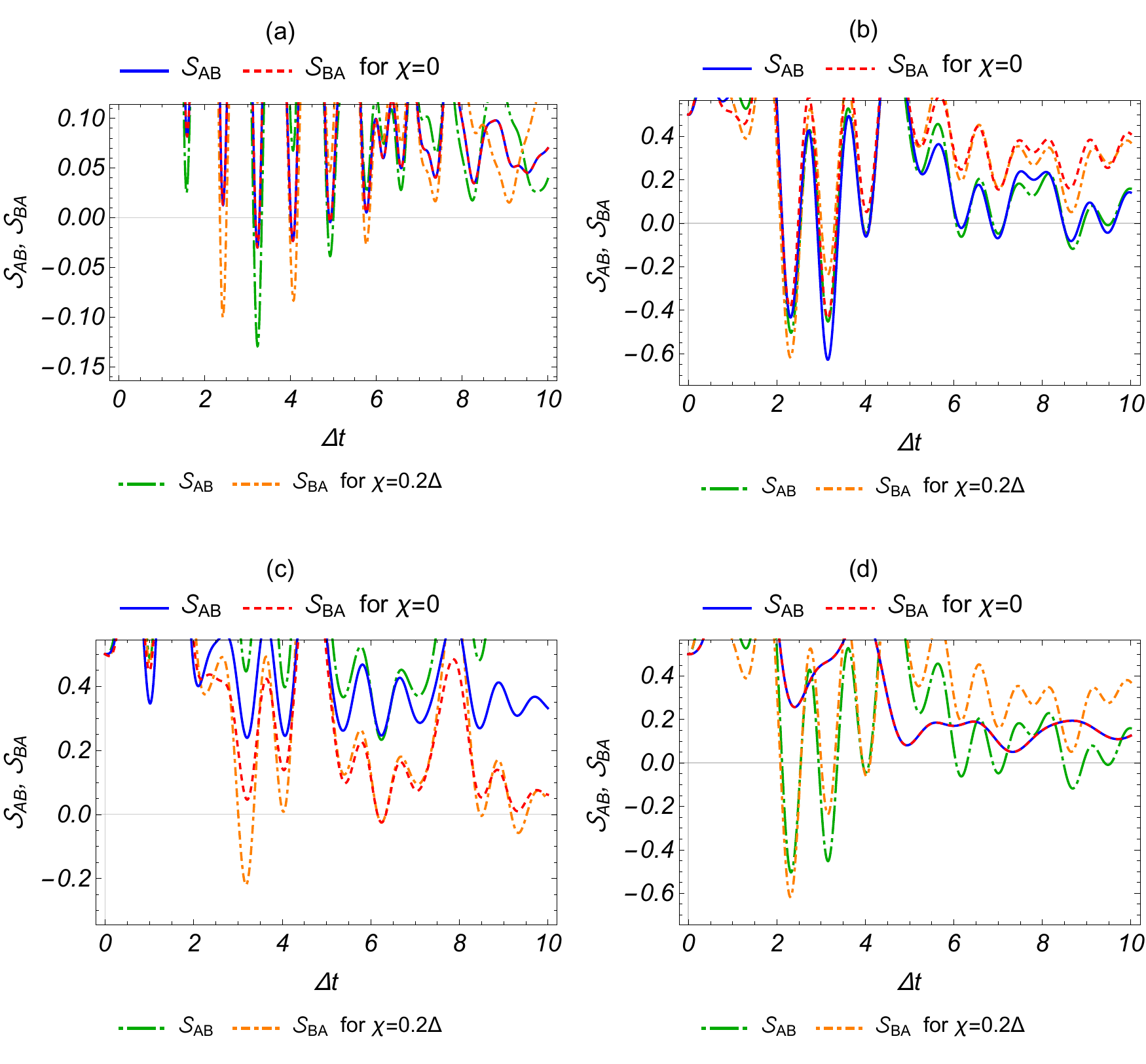}
	\caption{(Color online) Steering criteria as a function of $\Delta t$ for modes $\hat{A}$ and $\hat{B}$.  The figures (a), (b), (c) and (d) correspond to the AA, AN, NA and NN. Steering is confirmed if $\mathcal{S}_{\alpha \beta} < 0$, for modes $\alpha$ and $\beta$. }
	\label{STEER}
\end{figure*}

As we have already mentioned, entangled states may or may not satisfy steering condition, but a state satisfying steering condition must be entangled. Thus, a steering criterion can be viewed as a stronger criterion of non-classicality in comparison to the criteria of entanglement. Further, entangled states that are not steered cannot be used for one-way device independent quantum cryptography, but steered states can be \cite{branciard2012one}. This motivated us to look into the possibility of observing steered states in our system. To do so we have used the steering criterion given by Eq. (\ref{Steering}) and  plotted, for example, for two spatially separated ensembles modes $\hat{A}$ and $\hat{B}$ in  Fig.  (\ref{STEER}). The existence of the steered state  is observed for modes $\hat{A}$ and $\hat{B}$ (also for  $\hat{B}$ and $\hat{C}$ and for $\hat{A}$ and $\hat{C}$ as summarized in Table \ref{tab1}) for all configurations. Further, in contrast to entanglement witnesses, here we observe an  asymmetric nature of steering which  is reflected from Fig. \ref{STEER}, where it can be seen that $\mathcal{S}_{AB} \neq \mathcal{S}_{BA} $. For instance,  Fig. (\ref{STEER})-c shows that for non-zero driving field intensity, $\hat{A}$ can steer $\hat{B}$, while the converse is not observed. The results obtained for the steering criterion established the same observations as found for the Hillery Zubairy entanglement criteria. 
Failure to obtain steering in some of the cases (as highlighted in Table \ref{tab1}) establishes that it is a relatively stronger criterion of non-classicality, and the presence of steering correlations in two spatially separated  ensembles, mediated by the cavity field and controllable by the external driving, is an interesting observation.

\FloatBarrier	

\begin{figure}[ht] 
	\centering
	\includegraphics[width=70mm]{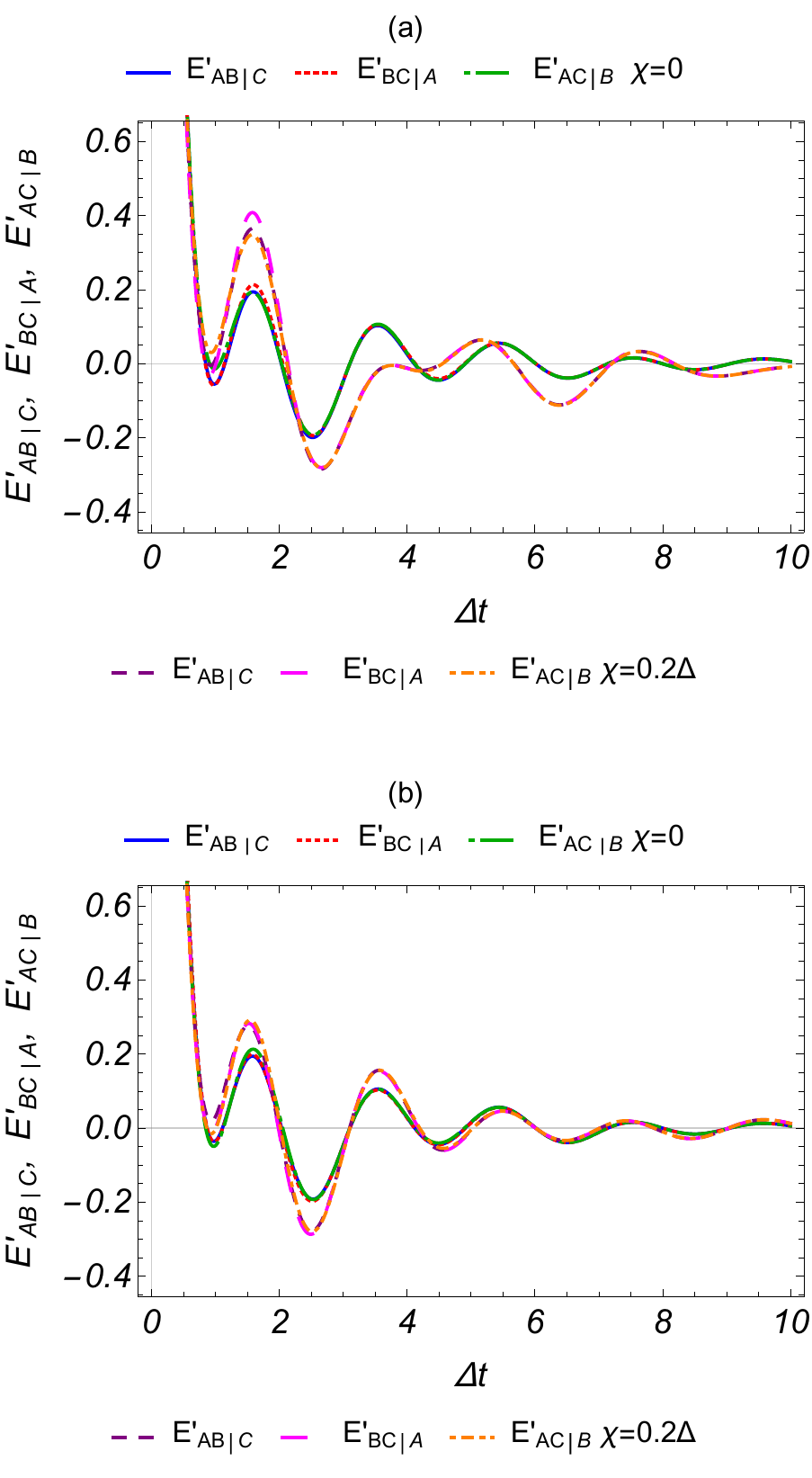}
	\caption{(Color online) Biseparability criteria as a function of $\Delta t$. Top and bottom panels correspond to AN and NA configurations, respectively. The nonseperability of modes $\alpha$, $\beta$ and $\gamma$ is implied by the satisfaction of either or both of the inequalities $E_{\alpha \beta|\gamma} < 0$ and  $E^{\prime}_{\alpha \beta|\gamma} <0$.}
	\label{Biseperability}
\end{figure}

So far we have seen quantum correlations involving two modes only. However, our system consists of three modes that are treated quantum mechanically. Hence, we may extend our study to check non-classical features of the system through some quantum correlations involving all the three modes $\hat{A}$, $\hat{B}$ and $\hat{C}$. To do so, use has been made of the biseperability criteria defined in Eqs. (\ref{Bisep1}) and (\ref{Bisep2}), to study multimode entanglement. Corresponding results are plotted in Fig. (\ref{Biseperability}). It is clear that the sufficient condition for the fully entangled state, which is the satisfaction of at least  one of the two sets of inequalities given by (\ref{set1}) and (\ref{set2}), is satisfied here. Thus, there exists a non-classical correlation involving all the three modes in both AN and NA configuration. Specifically, all possible combinations for the biseparability show almost similar variation in Fig. (\ref{Biseperability}). This establishes that the three-mode compound state is fully entangled. The application of the external field is found to enhance the signature for the existence of the multimode entanglement. Similar study for other configurations is summarized in Table \ref{tab1}.

\section{Conclusion}	 \label{Conclusion}
  In the previous sections, we have performed a detailed investigation on the temporal variation of various  witnesses of non-classicality present in a model physical system consisting of a cavity that contains  two ensembles of two level atoms which are placed in different configurations with respect to the Antinode and Node of the cavity field, viz., AA, AN, NA and NN configurations.  Further, it is considered that the left ensemble is driven by an external optical field which has been treated here as classical. The effects of this driving optical-field  on various witnesses of single mode (e.g., squeezing, Mandel's $Q$ parameter, antibunching) and intermodal (e.g., intermodal squeezing, antibunching, two and three mode entanglement, and steering) non-classicality have been studied systematically. The study has yielded various interesting results which are summarized in Table \ref{tab1}.
  
  	Before, we conclude the paper, we must note that one of the main findings of this paper is  that the optical-driving field may be used to control the amount of non-classicality. In fact, it can be used to enhance the non-classicality of the atomic ensemble modes, cavity modes and their compound modes.  Specifically, it is observed that the excitation mode $\hat{A}$ corresponding to the driven ensemble shows amplification in the witness of squeezing of  its quadratures in the presence of the external field ($\chi \neq 0$). Similar enhancement of the negative values of the non-classicality witness has also been observed  in the cavity photonic mode $\hat{C}$ when the driven ensemble is placed at the Antinode of the cavity field.
  	
  	Further, the existence of entanglement which is considered to be one of the main resources for quantum information processing has been observed here using a set of sufficient criteria for inseparability Specifically, we have used Hillery-Zubairy criterion and Duan's criterion for two-mode entanglement and a biseperability criteria for three mode scenario. Since steering can be used for one-way device independent quantum key distribution (\cite{branciard2012one} and references therein), we have also investigated the possibilities of observing steering involving various modes of the system. The investigation has not only revealed the existence of steering, but has also demonstrated its asymmetric nature.

The method adapted in this work is quite general and easy to follow. It can be extended easily to investigate the existence of non-classicality in similar physical systems of interest, specially in a set of driven cavity systems. For example, the study can be adapted to a system where both the ensembles are driven (with the same or different driving frequencies) or to a double cavity optomechanical system \cite{bai2016robust, yasir2016controlled, wang2016steady, li2016transparency, bariani2014hybrid, akram2015tunable}. In brief, non-classical features of various optomechanical, driven-  and non-driven cavity and optomechanics-like  systems can be studied using the technique used here. Further, the present system and similar driven-cavity systems can be treated in a completely quantum mechanical manner (by considering the driving optical field as weak and hence quantum mechanical) to reveal non-classicalities involving the mode(s) of the driving field(s). Such investigations are expected to yield various types of non-classicality in different physical systems that can be realized with the present technology and thus provide a wider choice of systems (to experimentalists) that can be used to build quantum devices that exploit the true power of the quantum world by producing and manipulating non-classical states. Keeping the above possibility in mind, we conclude this paper with an expectation that the present study will lead to a set of similar studies and subsequently to quantum devices having practical applications.
  	
\section*{Acknowledgments}

 AP thanks Department of Science and Technology (DST), India
for the support provided through the project number EMR/2015/000393.
KT acknowledges support from the Council of Scientific and Industrial
Research (CSIR), Government of India. The work of SB is supported by the project number 03(1369)/16/EMR-II funded by Council of Scientific and Industrial Research, New Delhi.

\section*{Appendix A: Equations of motion for involved operators}
\label{appendix:A}

\setcounter{equation}{0} \renewcommand{\theequation}{A.\arabic{equation}} 

In analogy of Eq. (\ref{eq:Lang-c}), we obtained Langevin equations for different single and compound modes as follows:

	\begin{flalign}
	\frac{d \avgA}{dt} = -i \Delta_a \avgA  -iG_A \avgc -i\chi - \frac{\Gamma_A}{2}\avgA,
	\label{eq:A} 
	\end{flalign}
	
	\begin{flalign}
	\frac{d \avgB}{dt} = -i\Delta_b \avgB  - iG_B \avgc - \frac{\Gamma_B}{2} \avgB,
	\label{eq:B}
	\end{flalign}
	
	\begin{align}
	\frac{d \avgc}{dt} &= -i \Delta_c \avgc  - iG_A \avgA - i G_B \avgB - \frac{\Gamma_c}{2} \avgc,
	\label{eq:c}
	\end{align}
	
	\begin{align}
	\frac{d \avgAA}{dt} &= -2i\Delta_a \avgAA -2i G_A \avgAc - 2i \chi \avgA  - \Gamma_a \avgAA,
	\label{AA} 
	\end{align}
	
	\begin{align}
	\frac{d \avgBB}{dt} &= -2i\Delta_b \avgBB  -2i G_B \avgBc - \Gamma_B \avgBB,
	\label{eq:BB} 
	\end{align}
	
	\begin{align}
	\frac{d \avgcc}{dt} &= -2i \Delta_c \avgcc  -2i G_A \avgAc -2i G_B \avgBc  - \Gamma_c \avgcc,
	\label{eq:cc} 
	\end{align}
	
	\begin{align}
	\frac{d \avgAdgr}{dt} &= i \Delta_a \avgAdgr  + i G_A \avgcdgr + i \chi - \frac{\Gamma_A}{2} \avgAdgr,
	\label{Adgr} 
	\end{align}
	
	\begin{align}
	\frac{d \avgBdgr}{dt} = i\Delta_b \avgBdgr  + i G_B \avgcdgr - \frac{\Gamma_B}{2} \avgBdgr,
	\label{eq:Bdgr}
	\end{align}
	
	\begin{align}
	\frac{d \avgcdgr}{dt} &= i \Delta_c \avgcdgr  	+ i G_A \avgAdgr + i G_B \avgBdgr - \frac{\Gamma_c}{2} \avgcdgr,
	\label{eq:cdgr}
	\end{align}
	
	\begin{align}\nonumber
	\frac{d \avgAdgrAdgr}{dt} &= 2i \Delta_a \avgAdgrAdgr  + 2iG_A \avgAdgrcdgr + 2i \chi \avgAdgr \\&- ~ \Gamma_A \avgAdgrAdgr,
	\label{eq:AdgrAdgr}
	\end{align}
	
	\begin{align}
	\frac{d \avgBdgrBdgr}{dt} = 2i \Delta_b \avgBdgrBdgr  + 2i G_B \avgBdgrcdgr - \Gamma_B \avgBdgrBdgr,
	\label{eqn:BdgrBdgr}
	\end{align}
	
	\begin{align}\nonumber
	\frac{d \avgcdgrcdgr}{dt} &= 2i \Delta_c \avgcdgrcdgr  + 2i G_A \avgAdgrcdgr + 2i G_B \avgBdgrcdgr \\&- \Gamma_c \avgcdgrcdgr,
	\label{eq:cdgrcdgr}
	\end{align}
	
	\begin{align}\nonumber
	\frac{d \avgAdgrA}{dt} &= i G_A[\avgAcdgr - \avgAdgrc]  + i\chi [\avgA - \avgAdgr] + \Gamma_A n_A \\& - \Gamma_A \avgAdgrA,
	\label{AdgrA} 
	\end{align}
	
	\begin{align}
	\frac{d \avgBdgrB}{dt} &= iG_B [\avgBcdgr - \avgBdgrc]  + \Gamma_{B} n_B - \Gamma_B \avgBdgrB,
	\label{BdgrB} 
	\end{align}
	
	\begin{align}\nonumber
	\frac{d \avgcdgrc }{dt} &= iG_A [\avgAdgrc - \avgAcdgr] + iG_B [\avgBdgrc - \avgBcdgr] \\&- \Gamma_c \avgcdgrc + \Gamma_c n_c,
	\label{cdgrc} 
	\end{align}
	
	\begin{align}\nonumber
	\frac{d\avgAB}{dt} &= -i(\Delta_a + \Delta_b) \avgAB   - iG_A \avgBc -iG_B \avgAc \\&- i \chi \avgB - \frac{\Gamma_A + \Gamma_B}{2} \avgAB,
	\label{eq:AB}
	\end{align}
	
	\begin{align}\nonumber
	\frac{d\avgABdgr}{dt} &= i[\Delta_b - \Delta_a] \avgABdgr  - iG_B \avgBdgrc + iG_B \avgAcdgr \\&- i\chi \avgBdgr - \frac{\Gamma_A + \Gamma_B}{2} \avgABdgr,
	\label{eq:ABdgr} 
	\end{align}
	
	\begin{align}\nonumber
	\frac{d\avgAdgrB}{dt} &= i(\Delta_a - \Delta_b) \avgAdgrB  + iG_A \avgBcdgr - iG_B \avgAdgrc \\&+ i \chi \avgB - \frac{\Gamma_A + \Gamma_B}{2} \avgAdgrB,
	\label{eq:AdgrB} 
	\end{align}
	
	\begin{align}\nonumber
	\frac{d \avgAdgrBdgr}{dt} &= i(\Delta_a + \Delta_b) \avgAdgrBdgr  + i G_A \avgBdgrcdgr + i G_B \avgAdgrcdgr \\&+ i\chi \avgBdgr - \frac{\Gamma_A + \Gamma_B}{2} \avgAdgrBdgr,
	\label{AdgrBdgr}
	\end{align}
	
	\begin{align}\nonumber
	\frac{d \avgBc}{dt} &= -i(\Delta_b + \Delta_c)\avgBc  - iG_B(\avgcc + \avgBB) \\&-iG_A \avgAB - \frac{\Gamma_B + \Gamma_c}{2} \avgBc,
	\label{Bc}
	\end{align}
	
	\begin{align}\nonumber
	\frac{d\avgBcdgr}{dt} &= -i(\Delta_b - \Delta_c) \avgBcdgr  + iG_B(\avgBdgrB - \avgcdgrc) \\&+ iG_A \avgAdgrB - \frac{\Gamma_B + \Gamma_c}{2}\avgBcdgr,
	\label{Bcdgr} 
	\end{align}
	
	\begin{align}\nonumber
	\frac{d\avgBdgrc}{dt} &= i(\Delta_b - \Delta_c)\avgBdgrc + iG_B [\avgcdgrc - \avgBdgrB] \\&- iG_A \avgABdgr  - \frac{\Gamma_B + \Gamma_c}{2} \avgBdgrc,
	\label{eq:Bdgrc} 
	\end{align}
	
	\begin{align}\nonumber
	\frac{d\avgBdgrcdgr}{dt} &= -i(\Delta_b + \Delta_c) \avgBdgrcdgr  + i G_B[\avgcdgrcdgr + \avgBdgrBdgr] \\&+ iG_A \avgAdgrBdgr -\frac{\Gamma_B + \Gamma_c}{2} \avgBdgrcdgr,
	\label{eq:Bdgrcdgr}
	\end{align}
	
	\begin{align}\nonumber
	\frac{d\avgAc}{dt} &= -i(\Delta_a + \Delta_c) \avgAc  - iG_A[\avgcc + \avgAA] -iG_B \avgAB  \\&-i\chi \avgc - \frac{\Gamma_A + \Gamma_c}{2} \avgAc,
	\label{Ac} 
	\end{align}
	
	\begin{align}\nonumber
	\frac{d\avgAcdgr}{dt} &= i(\Delta_c - \Delta_a) \avgAcdgr  + i G_A [\avgAdgrA - \avgcdgrc] \\&+ iG_B \avgABdgr -i\chi \avgcdgr - \frac{\Gamma_A + \Gamma_c}{2} \avgAcdgr,
	\label{Acdgr}
	\end{align}
	
	\begin{align}\nonumber
	\frac{d\avgAdgrc}{dt} &= i(\Delta_a - \Delta_c)\avgAdgrc + iG_A[\avgcdgrc -\avgAdgrA] \\&-iG_B \avgAdgrB + i\chi \avgc  -\frac{\Gamma_A + \Gamma_c}{2} \avgAdgrc,
	\label{Adgrc}
	\end{align}
	
	\begin{align}\nonumber
	\frac{d \avgAdgrcdgr}{dt} &= i(\Delta_a + \Delta_c) \avgAdgrcdgr  + iG_A[\avgcdgrcdgr + \avgAdgrAdgr] \\&+ iG_B \avgAdgrBdgr - \frac{\Gamma_A + \Gamma_c}{2} \avgAdgrcdgr.
	\label{Adgrcdgr}
	\end{align}
    Here, $n_A,\, n_B$ and $n_C$ represent the thermal photon numbers corresponding to mode $A,\, B$ and $C$, respectively.
    Also, it would be apt to note that one can express the various witnesses of non-classicality described in Section \ref{Criteria} in terms of the solutions of the above set of coupled differential equations, which can be obtained using Mathematica or similar programs or by using conventional methods of obtaining analytic solutions of the coupled differential equations. Particularly, in this work, we have used Mathematica to obtain simultaneous numerical solution of these coupled differential equations. 
    
    To illustrate the method adapted in this work to obtain the expressions for non-classicality witnesses using the solution of the above equations, we may consider the computation of  Mandel parameter as an example. Mandel parameter defined by Eq. (\ref{mandel}) contains the term $\langle (\hat{A}^\dagger \hat{A})^2 \rangle$. This quantity is not among the variables appearing in the above equations. So the solution of the above set of coupled equations would not provide us an expression for $\langle (\hat{A}^\dagger \hat{A})^2 \rangle$. To circumvent this problem, we have adapted a technique that allows us to simplify this term after writing it in normal-ordered form
    $\langle (\hat{A}^\dagger \hat{A})^2 \rangle = \langle \hat{A}^\dagger \hat{A}^\dagger \hat{A} \hat{A} \rangle + \langle \hat{A}^\dagger \hat{A} \rangle$
and subsequently using the decoupling relation \cite{anglin2001dynamics} $\langle ABCD \rangle \approx \langle AB \rangle \langle CD \rangle + \langle AC \rangle \langle BD  \rangle + \langle AD \rangle \langle BC \rangle - 2 \langle A \rangle \langle B \rangle \langle C \rangle \langle D \rangle$. Using this decoupling relation we can write 
    \begin{align}
\langle \hat{A}^\dagger \hat{A}^\dagger \hat{A} \hat{A} \rangle &\approx \langle \hat{A}^\dagger \hat{A}^\dagger  \rangle \langle \hat{A} \hat{A} \rangle + \langle \hat{A}^\dagger \hat{A} \rangle \langle \hat{A}^\dagger \hat{A}  \rangle \nonumber \\&+ \langle \hat{A}^\dagger \hat{A} \rangle \langle\hat{A}^\dagger \hat{A} \rangle  - 2 \langle \hat{A}^\dagger \rangle \langle \hat{A}^\dagger \rangle \langle \hat{A} \rangle \langle \hat{A} \rangle, \nonumber \\
                                                                &= \langle (\hat{A}^\dagger)^2 \rangle \langle \hat{A}^2 \rangle + 2 \langle   \hat{A}^\dagger \hat{A} \rangle^2 - 2\langle  \hat{A}^\dagger \rangle^2 \langle \hat{A} \rangle^2.
    \end{align}
Interestingly, the Mandel parameter can now be expressed in terms of the variables, time evolution of which is obtained as the solution of the above set of differential equation and we can express Mandel parameter (\ref{mandel parameter}) as
    \begin{equation}
    Q_M \approx \frac{\langle (\hat{A}^\dagger)^2 \rangle \langle \hat{A}^2 \rangle + \langle \hat{A}^\dagger \hat{A} \rangle^2 - 2\langle \hat{A}^\dagger \rangle^2 \langle \hat{A} \rangle^2   }{\langle   \hat{A}^\dagger \hat{A} \rangle}. 
        \end{equation} 

Clearly, the solution of the coupled differential equation listed above would now allow us to study the temporal evolution of the Mandel parameter and thus to investigate the presence of non-classicality in our system of interest under the framework of open quantum system. Similar strategy is adapted in the study of all other witnesses of non-classicality mentioned above and this is how the interesting results related to the temporal evolution of non-classicality witnesses illustrated in Figs. \ref{mandel}-\ref{Biseperability}, and summarized in Table \ref{tab1}, are obtained. 
\bibliographystyle{apsrev4-1}
\bibliography{References_1,biblio}

\clearpage
\widetext
\begin{center}
	\textbf{\large Supplemental Material: Probing non-classicality in an optically-driven  cavity with two atomic ensembles}
\end{center}
	We provide the various results which were obtained in the analysis carried out in our primary work. The model considered in this work consists of a cavity embedded with two ensembles of the two-level atoms.  It is found that the nonclassical features, demonstrated by the various witnesses, studied in this work, show enhancement when one of the ensembles in the cavity is driven by an external field. The system is studied in various configurations such as AA, AN, NA and NN as explained in our main paper. It is found that AA configuration is best suited for observing the nonclassical features in the model considered in this work.\par
\begin{figure*}[t]
	\centering
	\includegraphics[width=180mm, height=120mm]{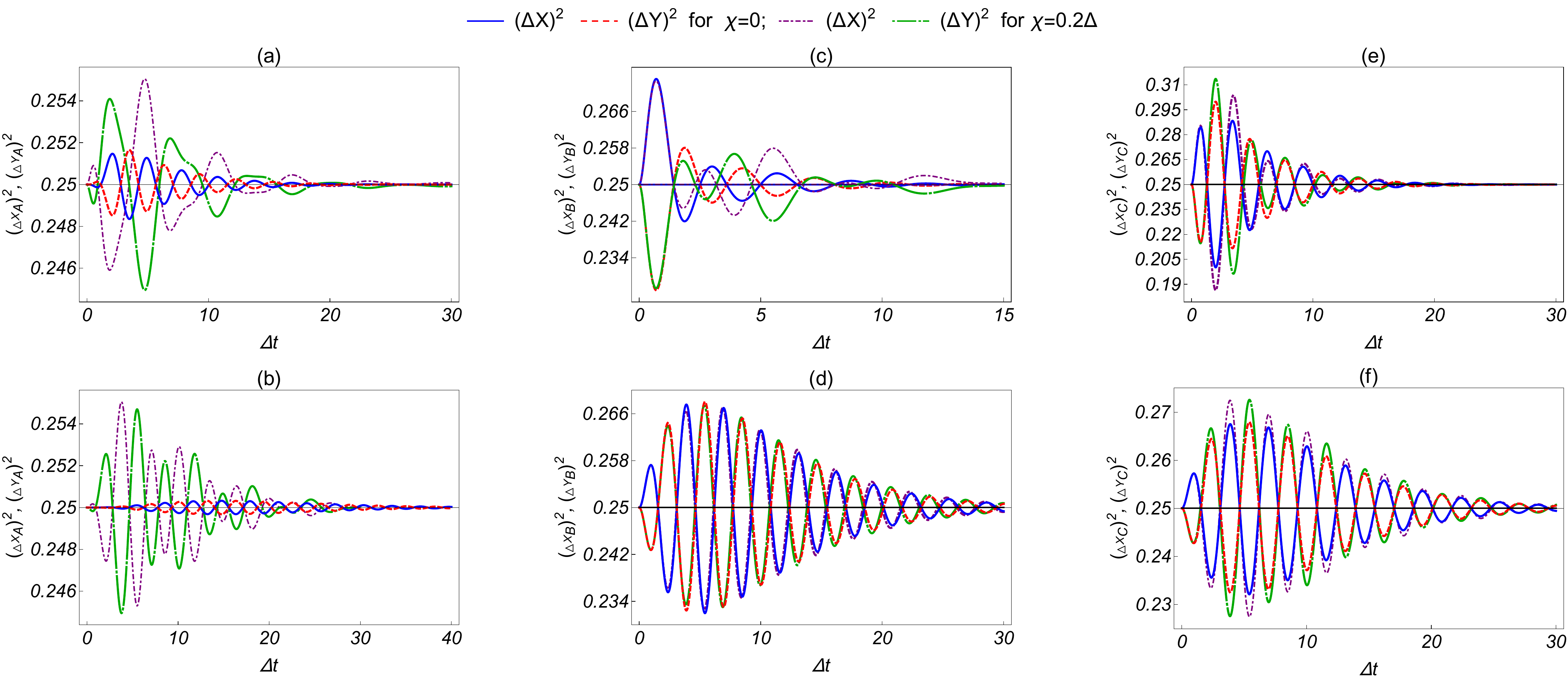}
	\caption{(Color online) Single mode squeezing plotted with respect to the dimensionless parameter $\Delta t$. Sub-figures (a)-(d), (b)-(e) and (c)-(f) correspond to modes $\hat{A}$, $\hat{B}$ and $\hat{C}$, respectively. The top and bottom panels pertain to the AA and NN configurations, respectively. It is clear that the application of the external field to the driven ensemble, that is, the non-zero value of $\chi$, enhances the squeezing in the respective quadratures of a particular mode.}
	\label{SingleModeSqueezing_AA_NN}
    \end{figure*}
\begin{figure*}[t]
	\centering
	\includegraphics[width=180mm]{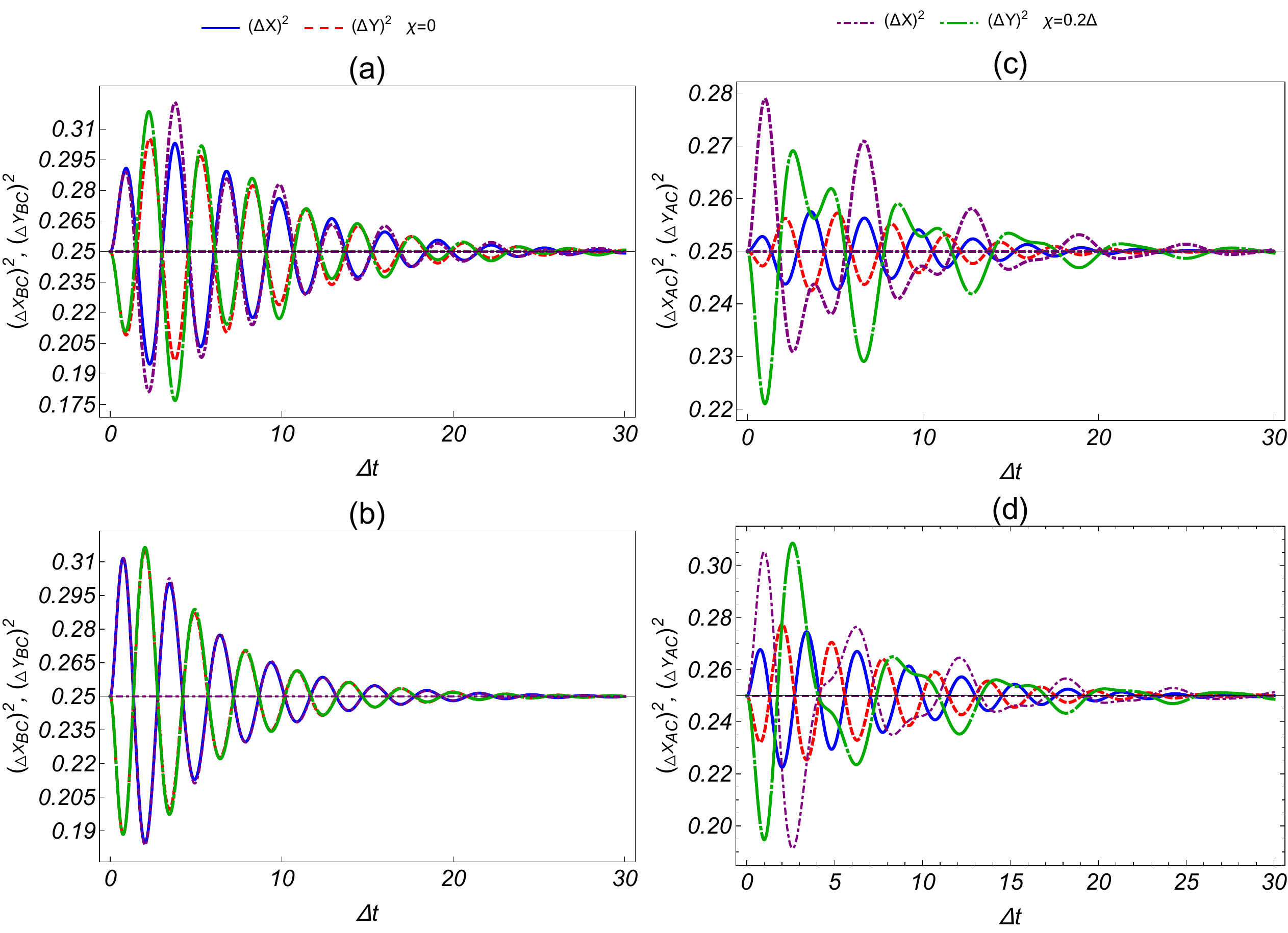}
	\caption{(Color online) Showing intermodal squeezing with respect to $\Delta t$. The Figs. (a) and (b) show the squeezing between modes $\hat{B}$ and $\hat{C}$. Figs. (c) and (d) show the same between modes $\hat{A}$ and $\hat{C}$.Top and bottom panels correspond to the AN and NA configurations, respectively. One finds enhancement in the intermodal squeezing as a result of driving ensemble ($A$) by the application of the external field.}
	\label{InterModalSqueezing_AN_NA}
    \end{figure*}
\begin{figure*}[t]
	\centering
	\includegraphics[width=180mm, height=120mm]{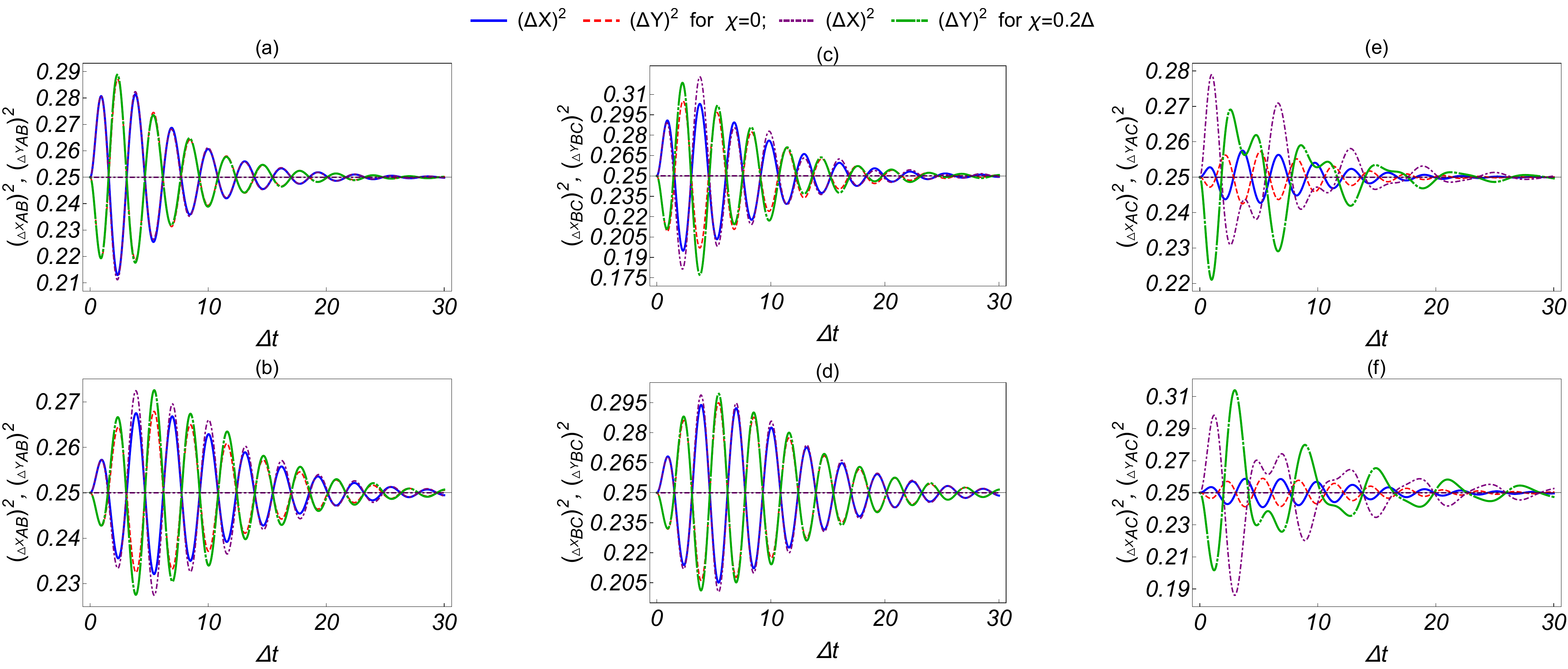}
	\caption{(Color online) Intermodal squeezing plotted with respect to $\Delta t$. Sub-figures (a)-(d) show the squeezing with modes $\hat{A}$ and $\hat{B}$ , (b)-(e) with modes $\hat{B}$ and $\hat{C}$ and (c)-(f) with modes $\hat{A}$ and $\hat{C}$. The top and bottom panels pertain to the AA and NN configurations, respectively.}
	\label{InterModalSqueezing_AA_NN}
    \end{figure*}
\begin{figure*}[ht]
	\centering
	\includegraphics[width=0.75\linewidth]{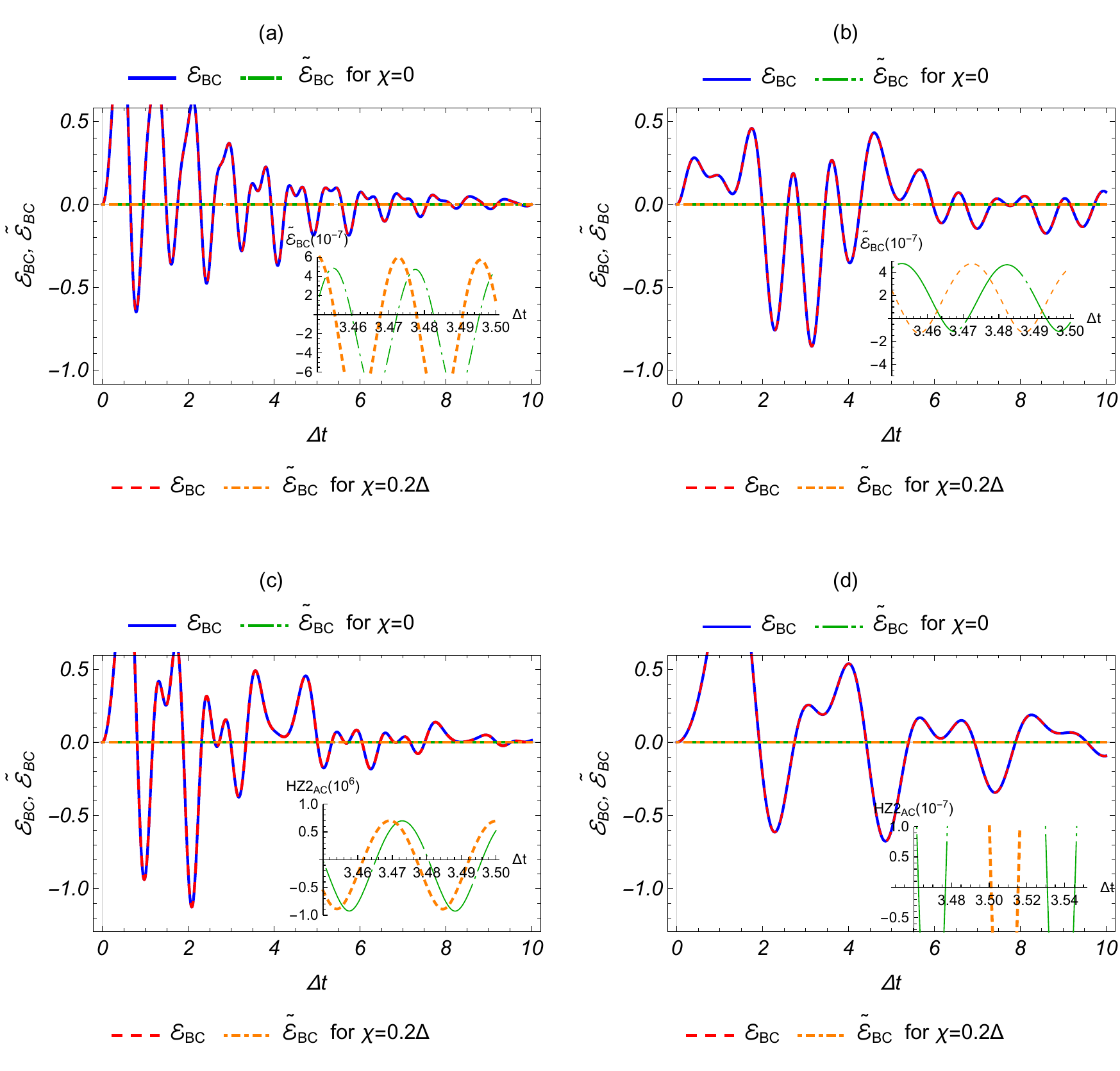}
	\caption{(Color online) Hillery Zubairy  criteria as a function of $\Delta t$ for modes $\hat{B}$ and $\hat{C}$. The figures (a), (b), (c) and (d) correspond to AA, AN, NA and NN  configurations, respectively. Entanglement is confirmed if $\mathcal{E}_{\alpha \beta} < 0$, for modes $\alpha$ and $\beta$. }
	\label{HZ_BC} 
    \end{figure*}
\begin{figure*}[ht]
	\centering
	\includegraphics[width=0.75\linewidth]{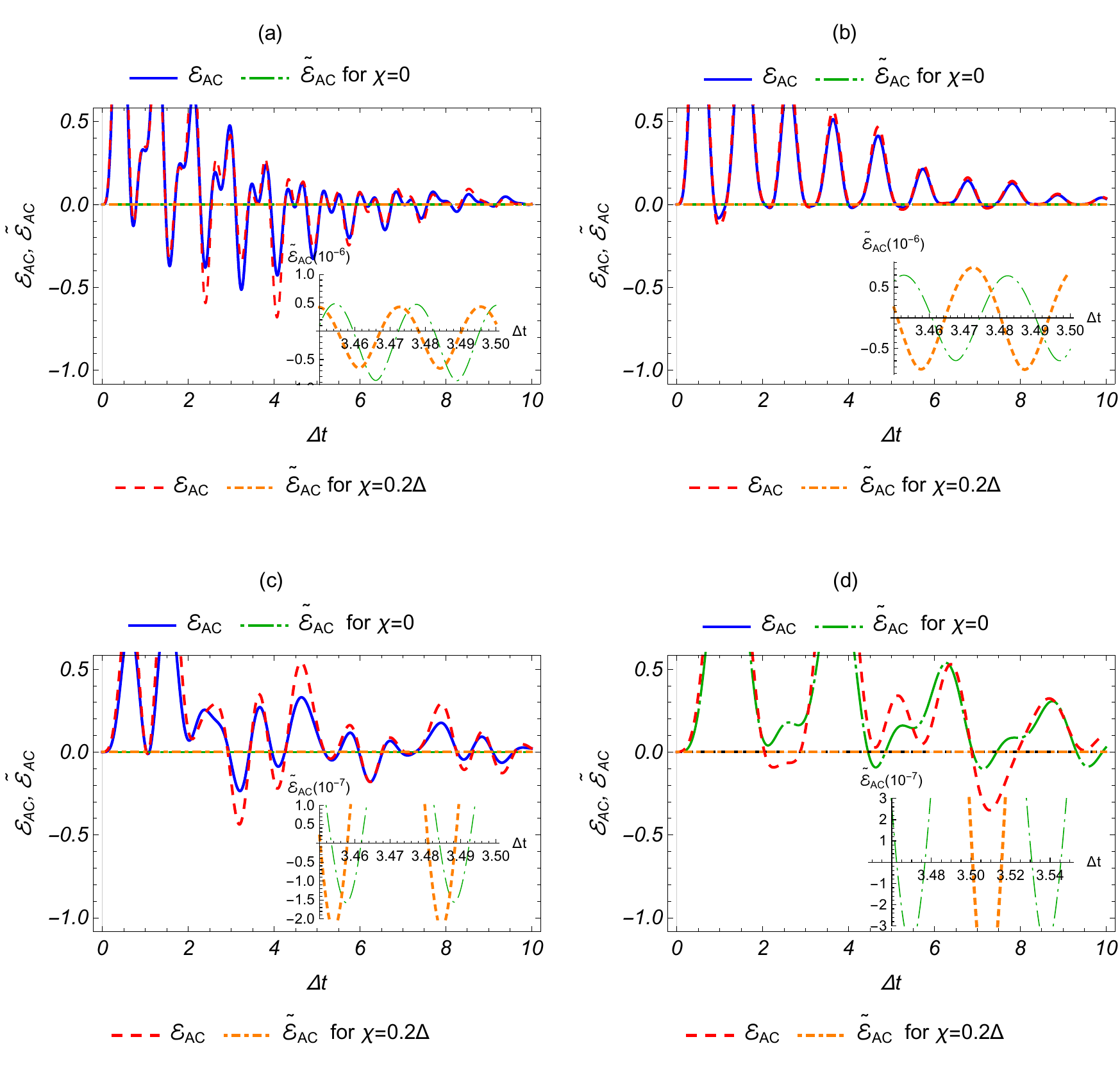}
	\caption{(Color online) Hillery Zubairy  criteria as a function of $\Delta t$ for modes $\hat{A}$ and $\hat{C}$. The figures (a), (b), (c) and (d) correspond to AA, AN, NA and NN  configurations, respectively. Entanglement is confirmed if $\mathcal{E}_{\alpha \beta} < 0$, for modes $\alpha$ and $\beta$. }
	\label{HZ_AC} 
    \end{figure*}
\begin{figure}[ht] 
	\centering
	\includegraphics[width=70mm]{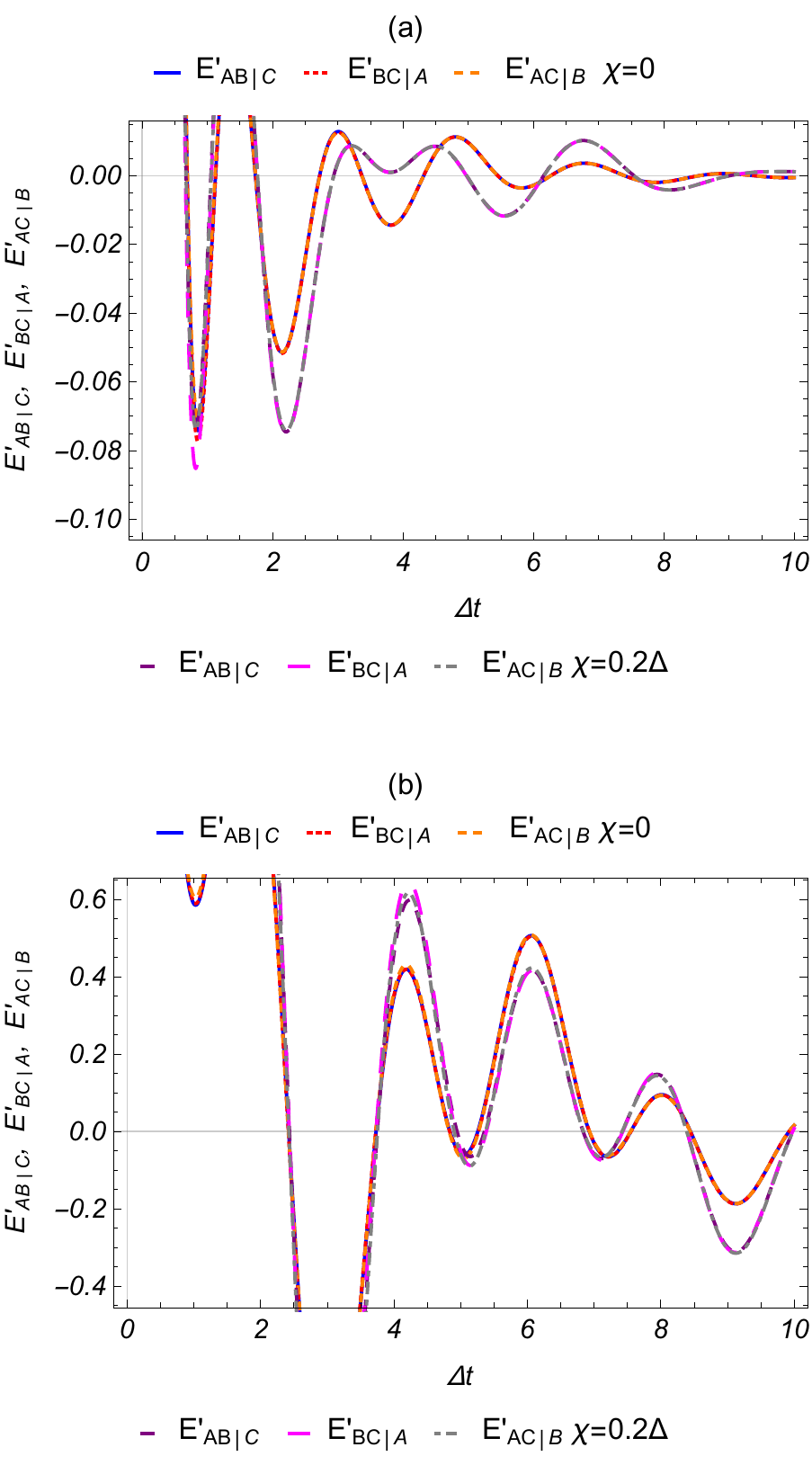}
	\caption{(Color online) Showing biseperability criteria plotted against $\Delta t$. Figs. (a) and (b) correspond to AA and NN configurations, respectively.}
	\label{Bisep_AA_NN}
    \end{figure}
\begin{figure*}[ht] 
	\centering
	\begin{tabular}{cc}
	\includegraphics[width=0.5\linewidth]{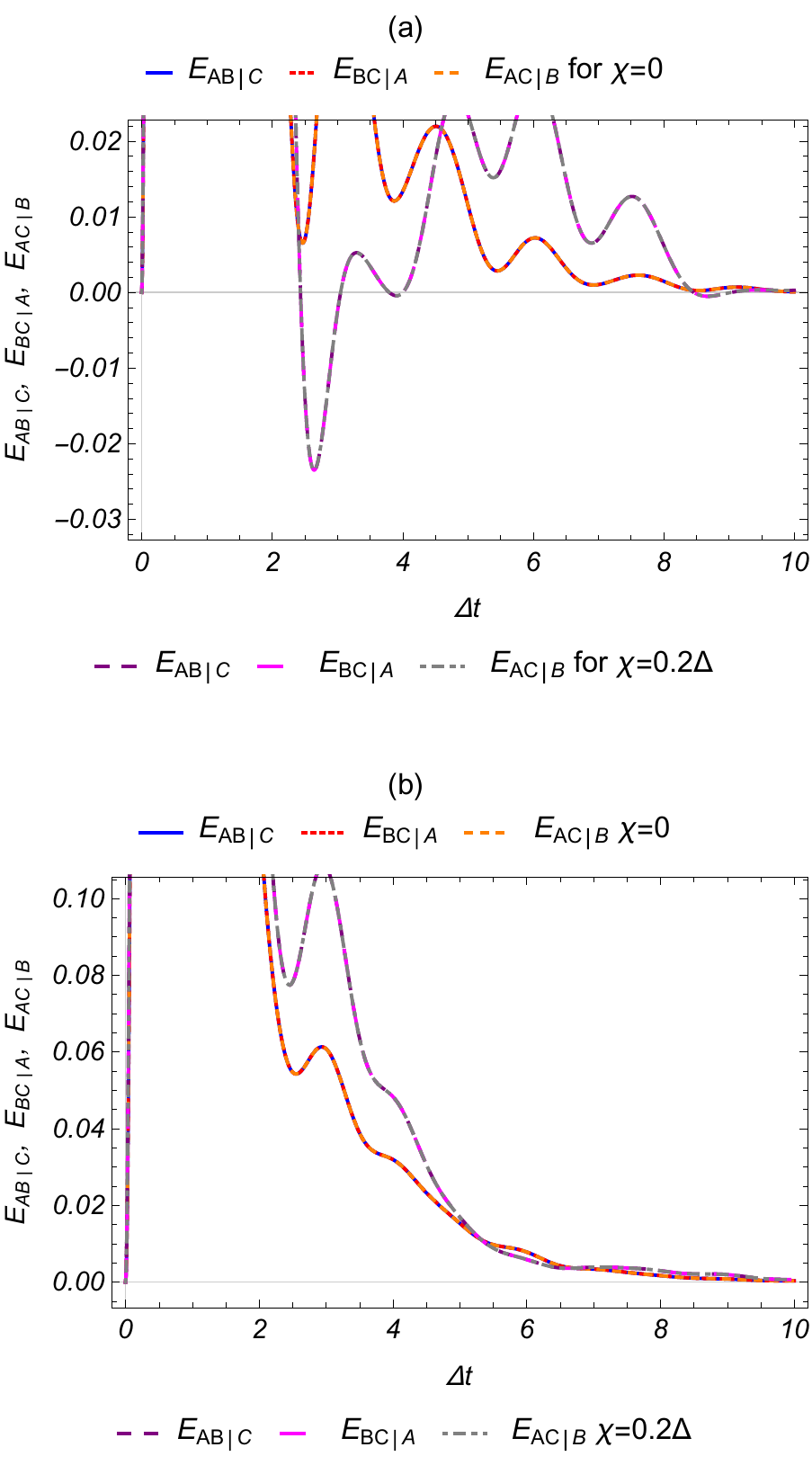}
	\includegraphics[width=0.5\linewidth]{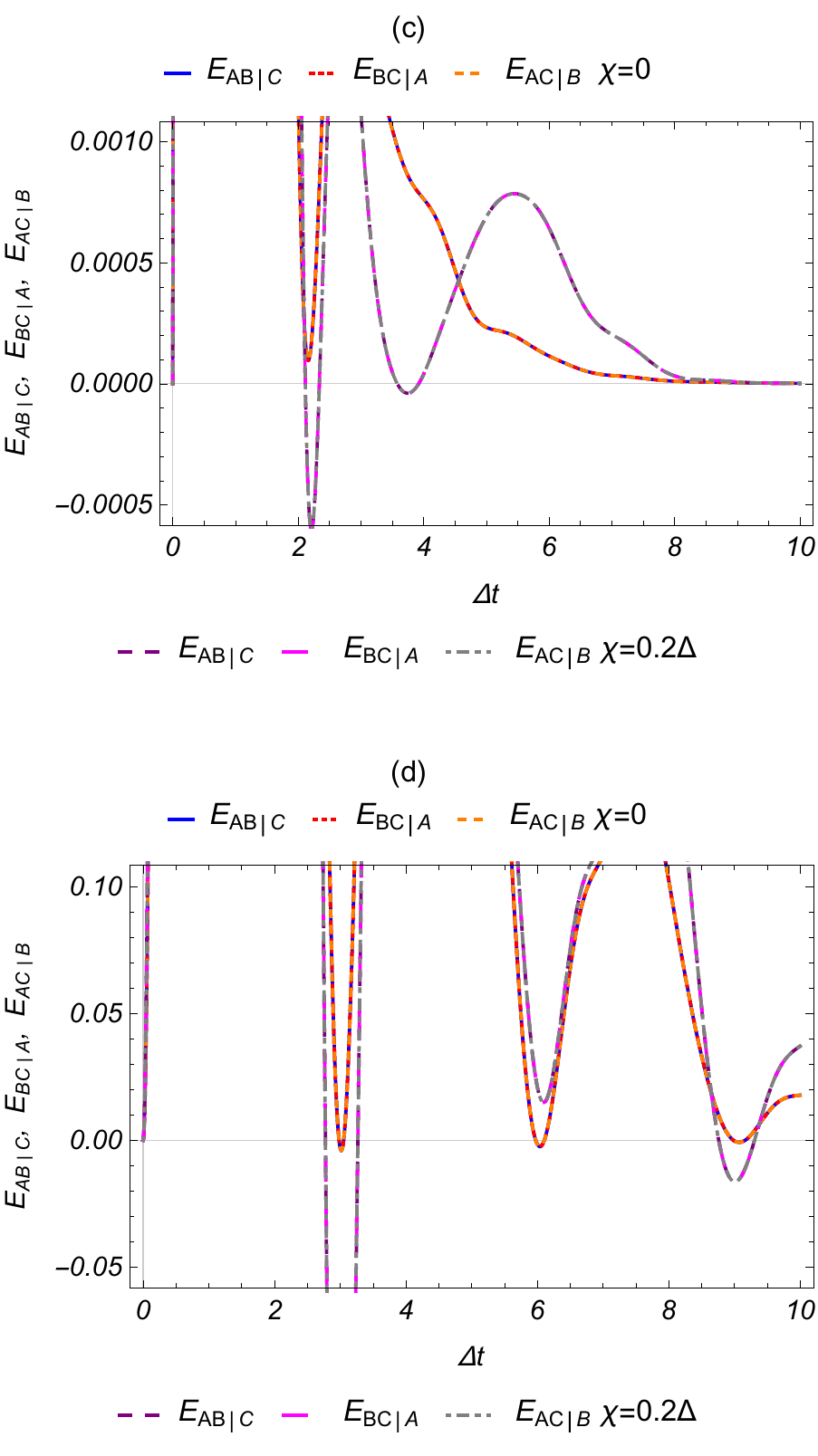}
	\end{tabular}
	\caption{(Color online) Biseperability criterion. (a) AN configuration, (b) NA configuration, (c) AA configuration, (d) NN configuration. }
	\label{Bisep_E_AN_NA_AA_NN} 
    \end{figure*}
\begin{figure*}[t]
	\centering
	\includegraphics[width=0.90\linewidth]{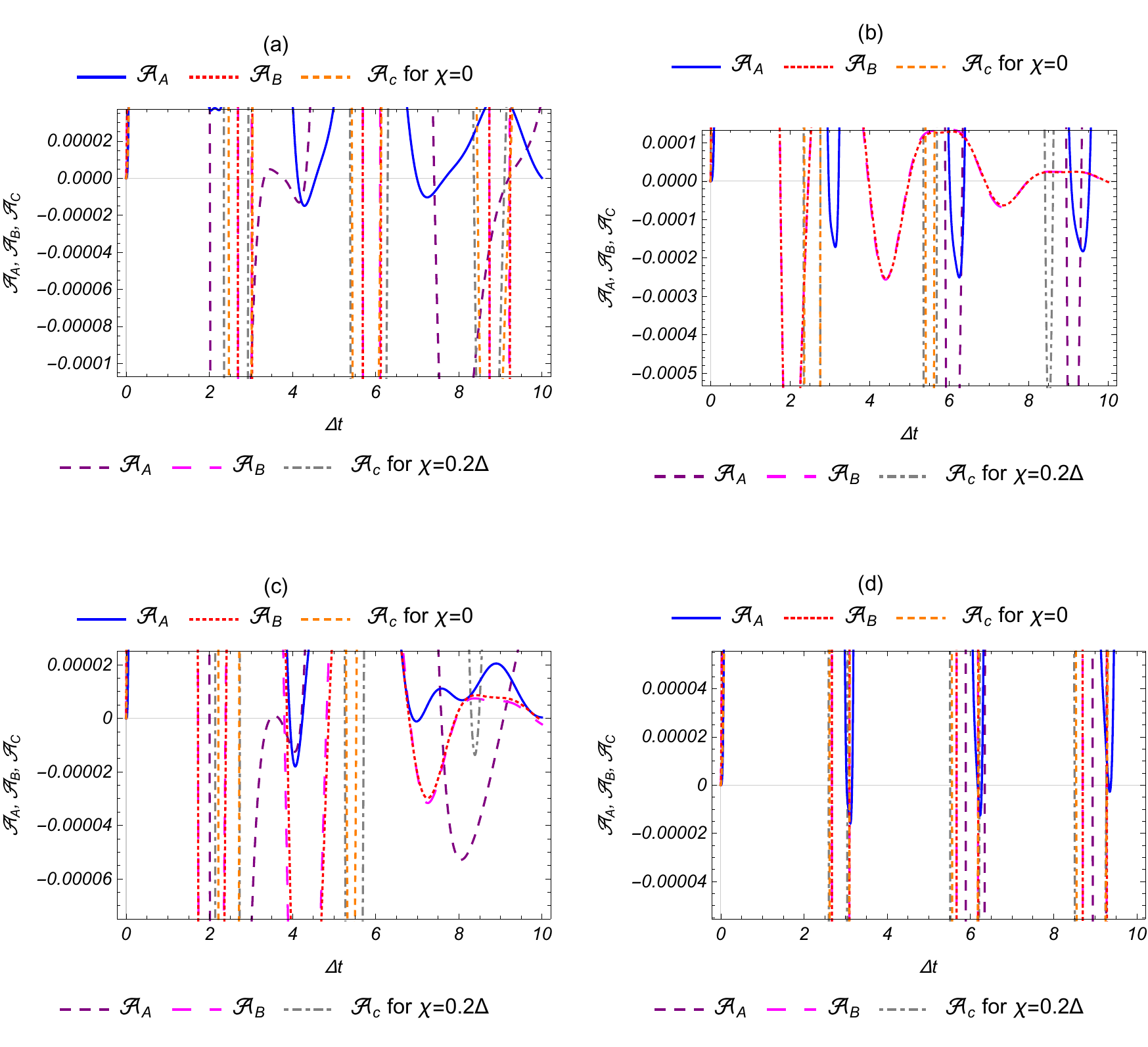} 
	\caption{(Color online) Single-mode antibunching. (a) AN configuration, (b) NA configuration, (c) AA configuration, (d) NN configuration. }
	\label{SingleMode_Antibunching} 
    \end{figure*}
\begin{figure*}[ht] 
	\centering
	\includegraphics[width=130mm]{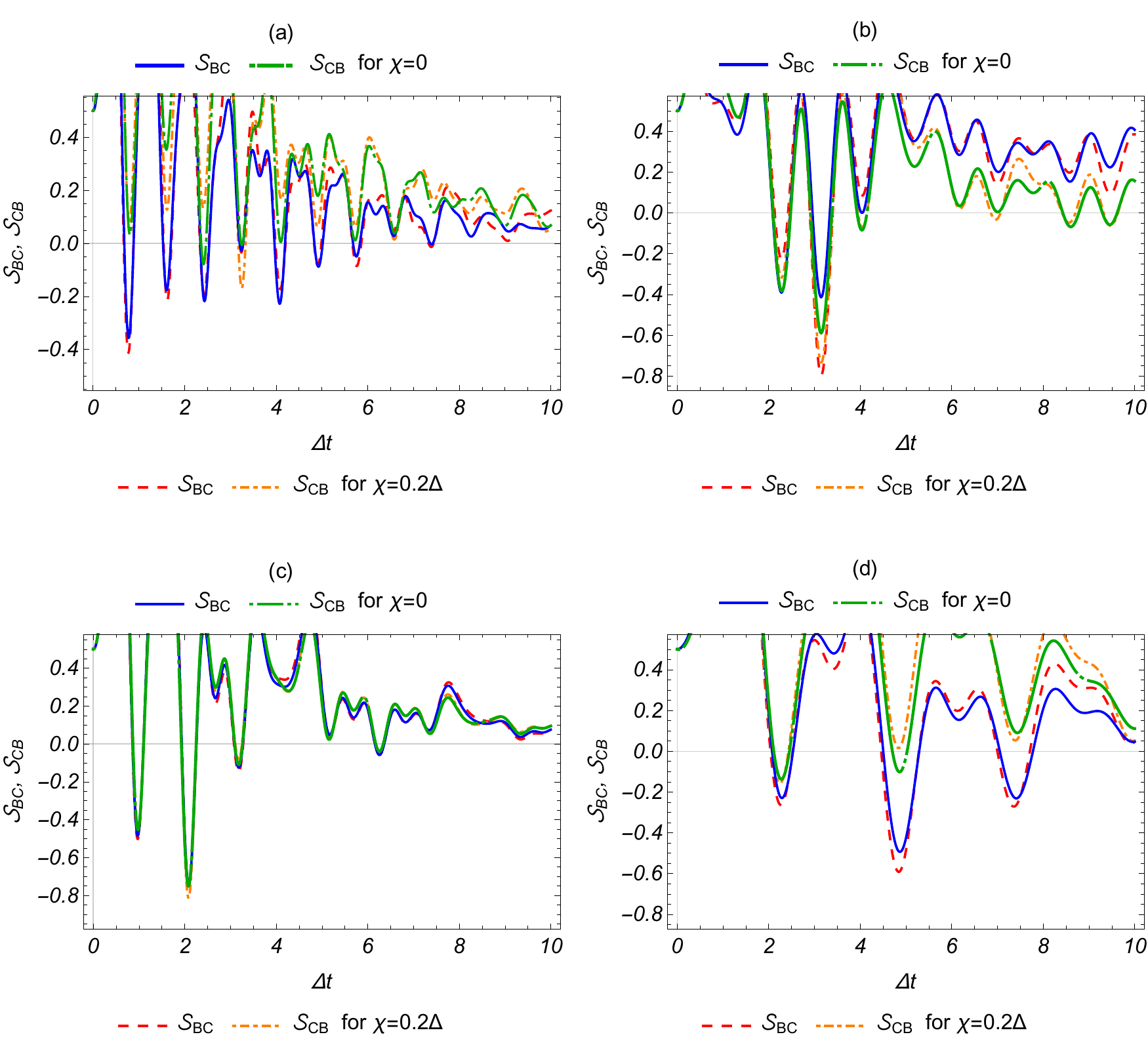}
	\caption{(Color online) Steering criterion for modes $\hat{B}$ and $\hat{C}$, plotted with respect to the parameter  $\Delta t$. Figs. (a), (b), (c) and (d) correspond to AA, AN, NA and NN configurations, respectively.}
	\label{Steering_BC}
    \end{figure*}
\begin{figure*}[ht] 
	\centering
	\includegraphics[width=130mm]{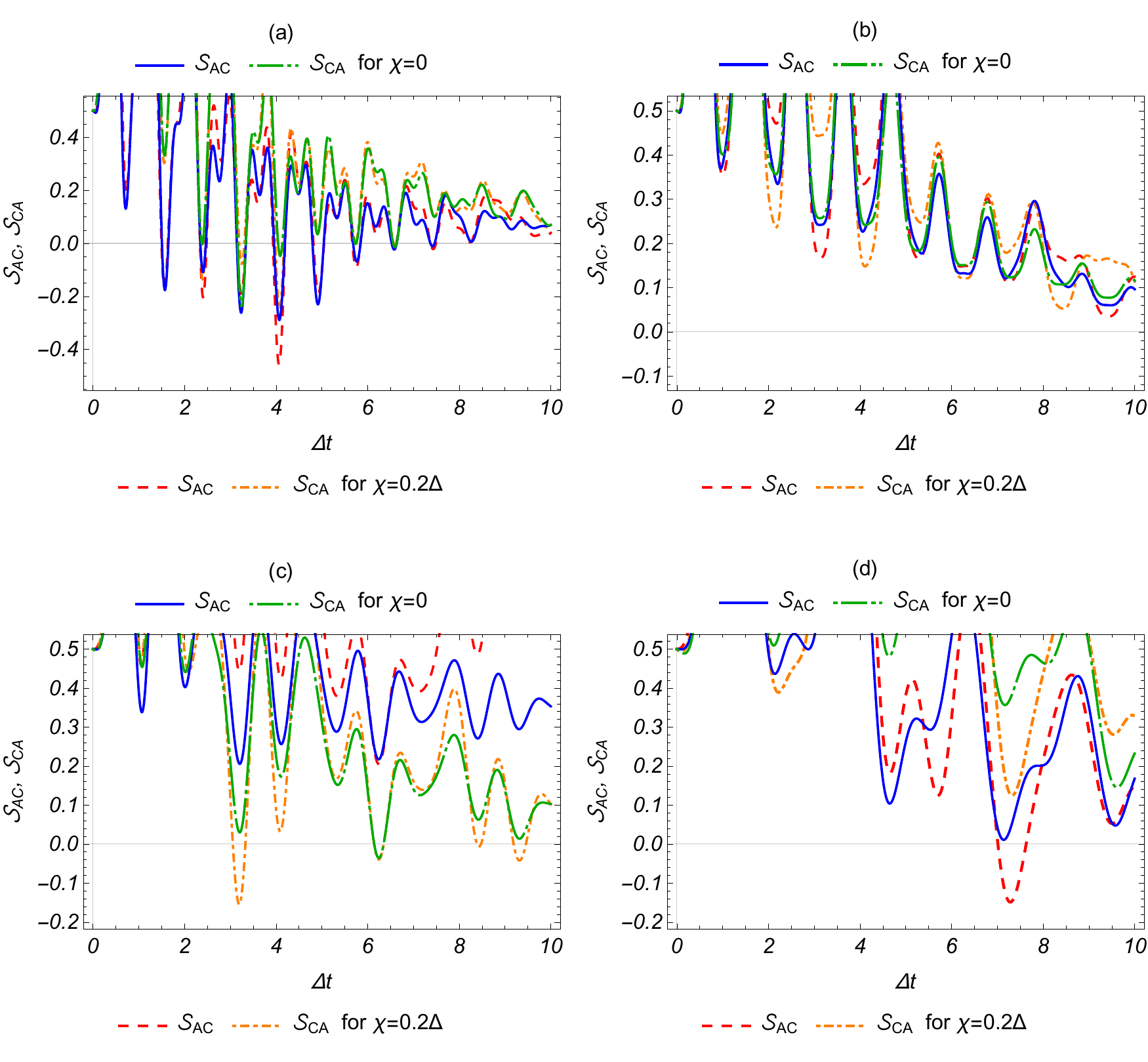}
	\caption{(Color online) Steering criterion for modes $\hat{A}$ and $\hat{C}$, plotted with respect to the parameter  $\Delta t$. Figs. (a), (b), (c) and (d) correspond to AA, AN, NA and NN configurations, respectively.}
	\label{Steering_AC}
    \end{figure*}

	In what follows, we have shown the behavior of the nonclassicality witnesses in the cases not displayed in the main paper but included in Table (I). Specifically, time evolution of number operators corresponding to each mode in AA and NN configurations is shown in Fig. (\ref{number}). A similar study for Mandal parameter (Fig. (\ref{mandel})), Duan et al.'s criterion  (Fig. (\ref{duan})), single-mode squeezing (Fig.  (\ref{SingleModeSqueezing_AA_NN})), intermodal squeezing (Figs. (\ref{InterModalSqueezing_AN_NA}) and (\ref{InterModalSqueezing_AA_NN})), Hillery-Zubariy criteria (Figs. (\ref{HZ_BC}) and (\ref{HZ_AC})), intermodal antibunching (Fig.(\ref{IMAB})), Biseperability criteria (Figs. (\ref{Bisep_AA_NN}) and (\ref{Bisep_E_AN_NA_AA_NN})), single-mode antibunching (Fig. (\ref{SingleMode_Antibunching})) and steering (Figs. (\ref{Steering_BC}) and (\ref{Steering_AC})). All the quantities are plotted with respect to the dimensionless parameter $\Delta t$, where $\Delta$ is the  common detuning.
	
	\begin{figure}[ht] 
	\centering
	\includegraphics[width=70mm]{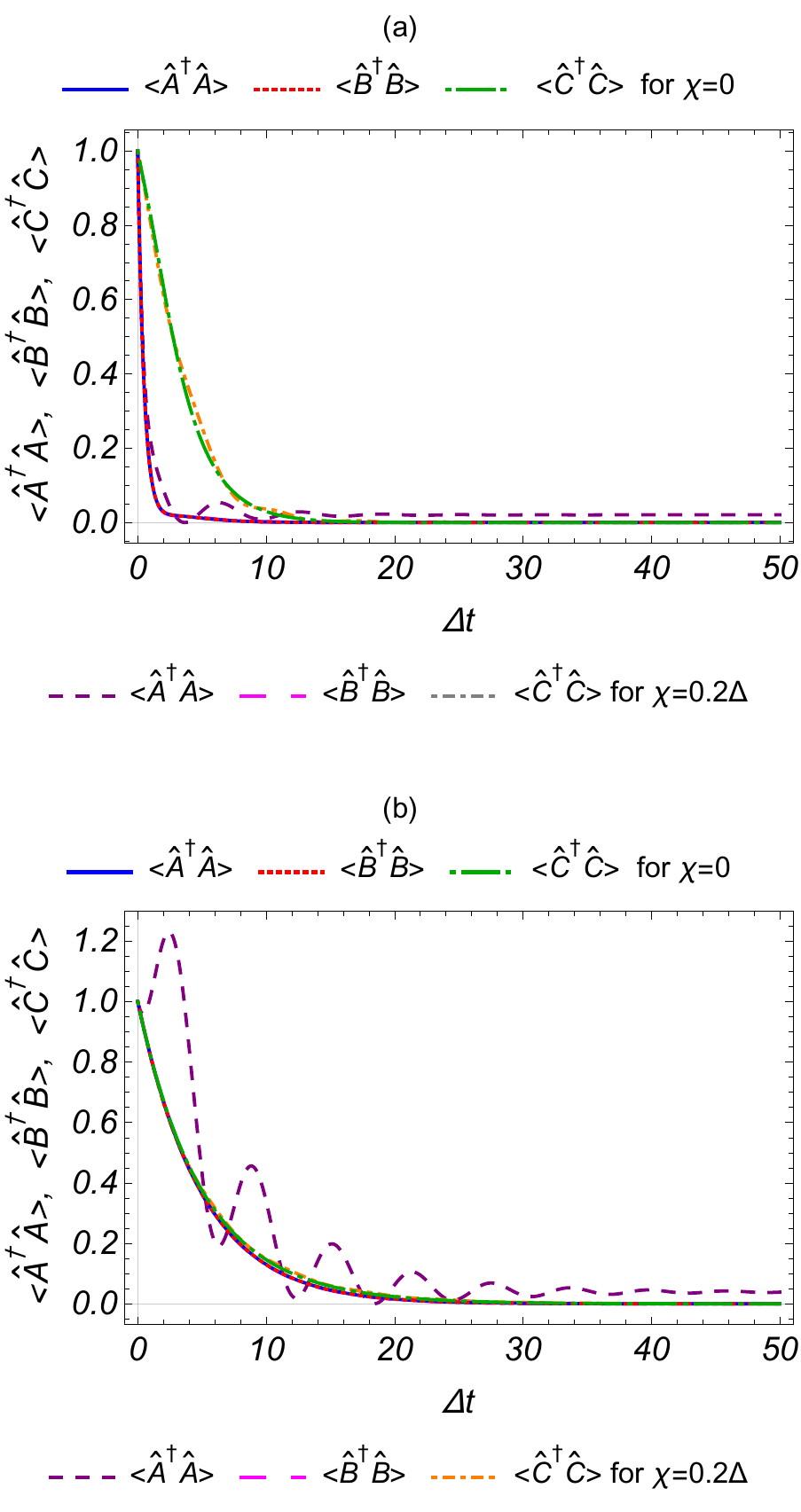}
	\caption{(Color online) Average number of cavity photons and excitations corresponding to the two ensembles, studied with respect to the dimensionless parameter $\Delta t$. Figs. (a) and (b)  correspond to AA (both the ensembles placed at Antinode) and NN (both the ensembles  placed at Node) configurations, respectively. The average number of excitations corresponding to the driven ensemble ($\langle A^{\dagger} A \rangle$), the undriven ensemble ($\langle B^{\dagger} B \rangle$) and the average cavity photon number ($\langle c^{\dagger} c \rangle$) is depicted for $\chi = 0$ and $\chi = 0.2 \Delta$.}
	\label{number}
	\end{figure}

	\begin{figure}[ht] 
	\centering
	\includegraphics[width=70mm]{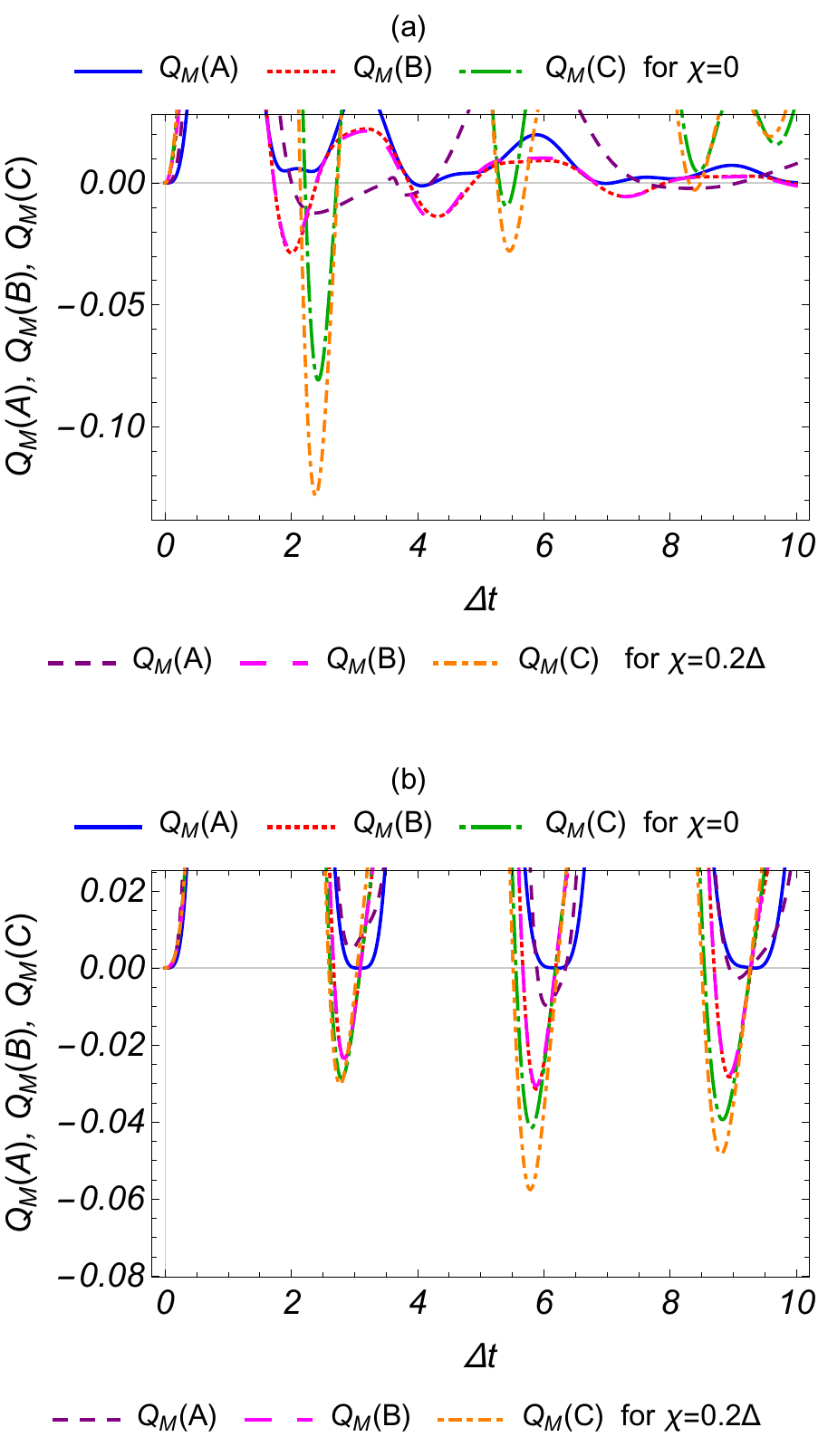}
	\caption{(Color online) Mandel parameter with respect to the dimensionless parameter $\Delta t$. Figs. (a) and (b) correspond to AA and NN configurations, respectively. The nonclassical nature of the field corresponding to the mode $\alpha$  is confirmed by $Q_M(\alpha) < 0$.}
		\label{mandel}
	\end{figure}

\begin{figure}[ht] 
	\centering
	\includegraphics[width=70mm]{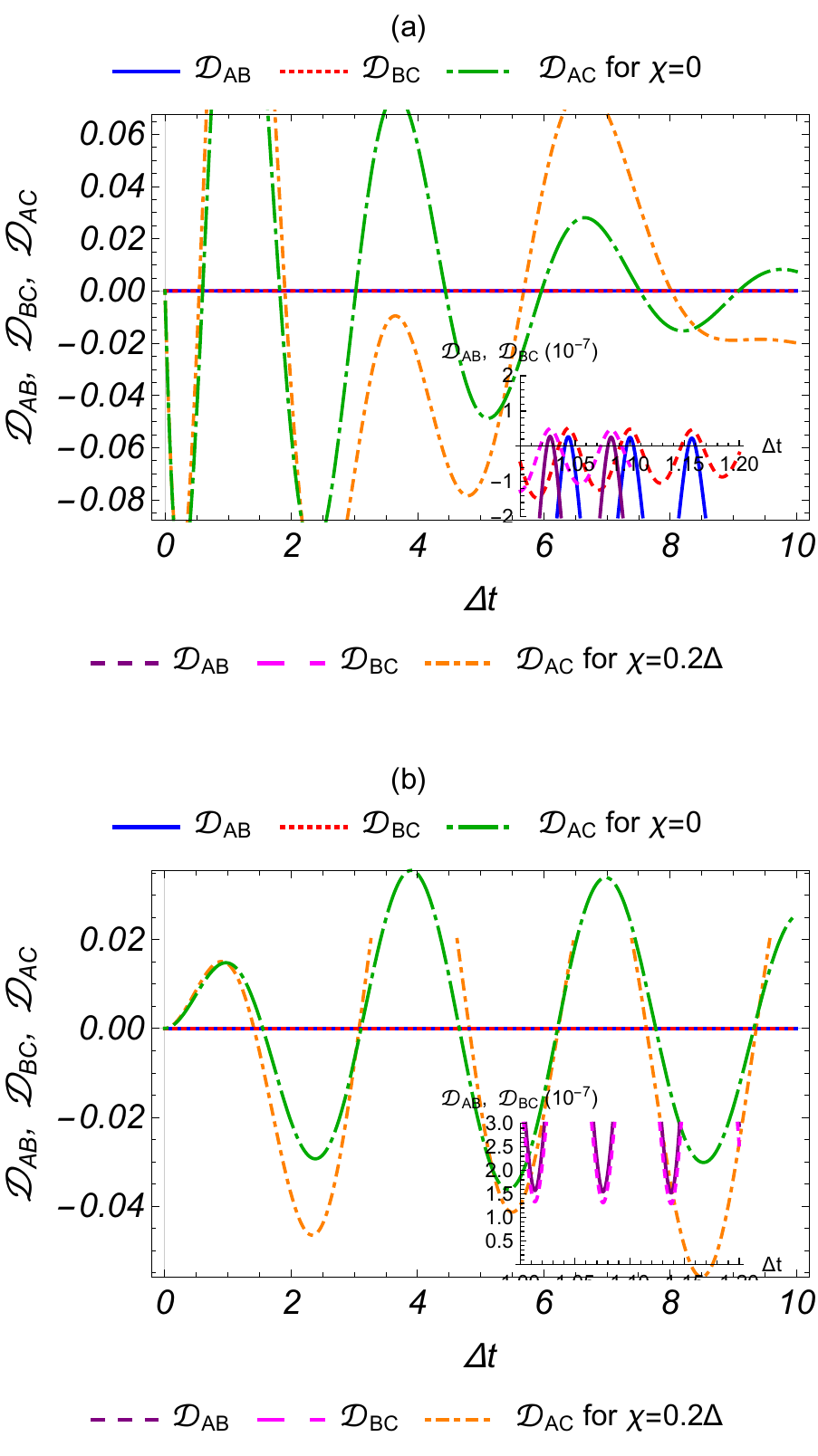}
	\caption{(Color online) Showing Duan separability parameter against $\Delta t$. Figs. (a) and (b) correspond to AA and NN configurations, respectively. The sufficient condition for inseparability is implied by $\mathcal{D}_{\alpha \beta} < 0$, for modes $\alpha$ and $\beta$.}
	\label{duan}
\end{figure}

    \begin{figure}[ht] 
    \centering
    \includegraphics[width=70mm]{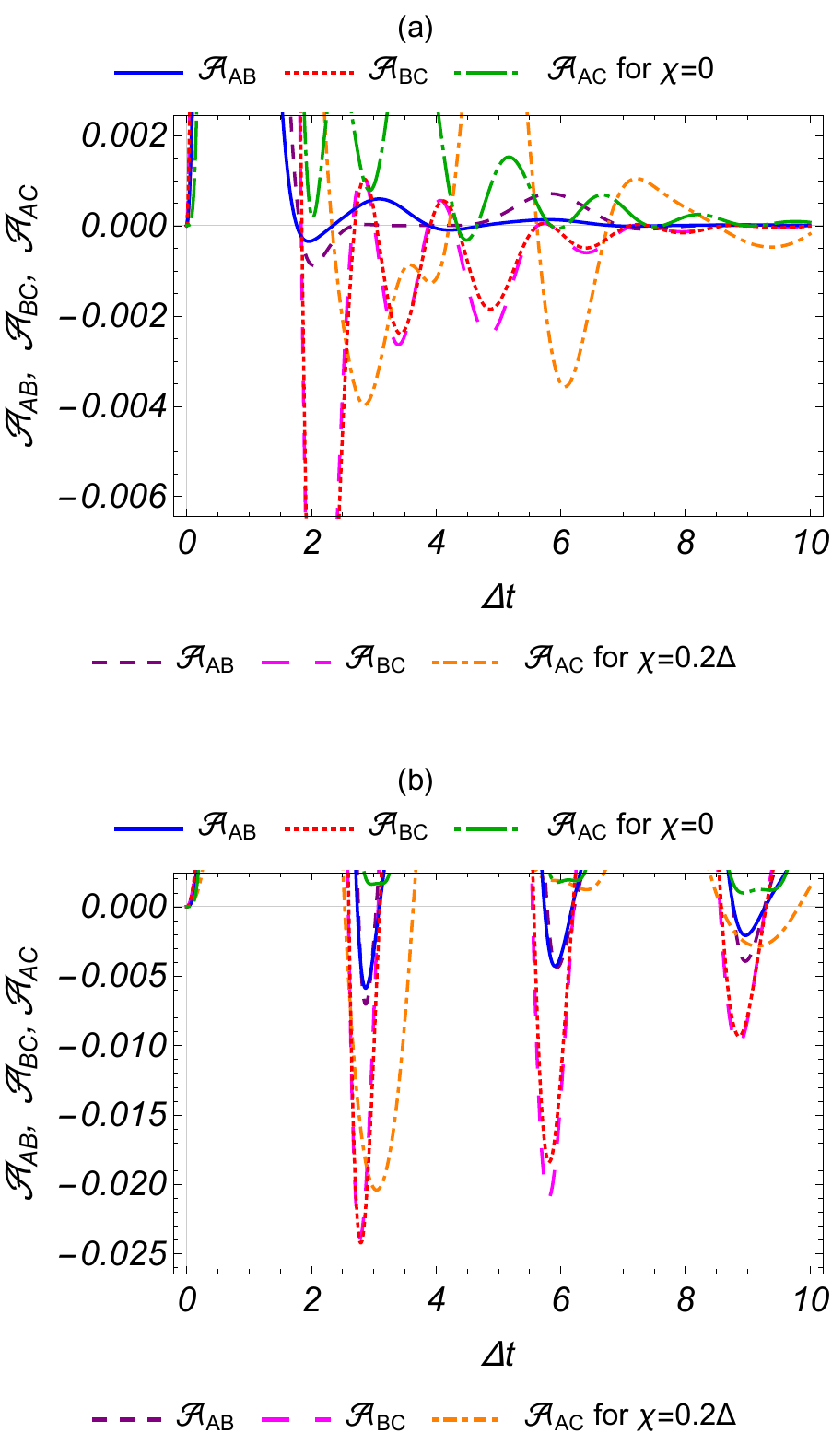}
    \caption{(Color online) Showing intermodal anti-bunching parameter $\mathcal{A}_{\alpha \beta}$ for modes $\alpha$ and $\beta$, plotted with respect to the parameter  $\Delta t$. Top (a) and bottom (b) panels correspond to AA and NN configurations, respectively. The existence of intermodal antibunching is confirmed if $\mathcal{A}_{\alpha \beta} < 0$. }
    \label{IMAB}
    \end{figure}

\end{document}